\documentclass[twocolumn,showpacs,preprintnumbers,amsmath,amssymb]{revtex4}

\usepackage{graphicx}
\usepackage{dcolumn}
\usepackage{bm}

\begin{document}

\preprint{APS/123-QED}
\title{Dilaton Dynamics in ${\rm (A)dS_5\times S^5}$}

\author{Kunihito Uzawa}
\address{Graduate School of Human and Environmental Studies,\\
Kyoto University, Kyoto 606-8501, Japan}%

\author{Kei-ichi Maeda}
\address{Department of Physics, Waseda University, 
Tokyo 169-8555, Japan\\
Waseda Institute for Astrophysics, Waseda University, Tokyo 169-8555,
Japan.\\
\\
Advanced Research Institute for Science and
Engineering, Waseda University, Tokyo 169-8555, Japan.\\
\\
}%

\date{\today}


\begin{abstract} 
We investigate a stabilization of extra dimensions in a ten
 dimensional Kaluza-Klein theory and IIB supergravity.
We assume $(A)dS_5\times S^5$ compactifcation,
and calculate quantum effects to find an effective potential for the 
radius of internal space.
The effective potential has a minimum,
and if the universe is created on the top of the potential hill,
the universe evolves from $dS_5$ to $AdS_5$ after exponential expansion.
The  internal space $S^5$ stays to be small and its radius becomes
constant.
 Our model in IIB supergravity 
contains the 4-form gauge field
 with classically vacuum expectation value, which is role of ten
 dimensional cosmological constant.
If the universe evolves into AdS,
 the five dimensional Randall \& Sundrum setup with
 stabilized dilaton is obtained from
 the type IIB supergravity model.
\end{abstract}

\pacs{Valid PACS appear here}

\maketitle



\section{Introduction}
	\label{sec:introduction}
Randall \& Sundrum (RS) brane world model\cite{ra}
has been investigated by
many authors in the cosmological and gravitational points of view.  
This model shows several interesting properties. 
The hierarchy problem may be solved by the warp factor.
Or the gravity can be confined in non-compact spacetime.
However, a higher dimensional realistic model including 
RS setup is not  so far known. The most plausible candidate is
the $AdS_5\times S^5$ compactification of 
the ten dimensional type IIB supergravity theory\cite{du}.   
The scale of $S^5$, which is expected to be much smaller than 
that of $AdS_5$, 
 is determined by the dilaton arising from a
spontaneous compactification and is generally dynamical variable in the
cosmological point of view. 
The dilaton in type IIB supergravity
 model has not ever been discussed as a dynamical variable but 
assumed to be a constant
parameter.
In order to discuss whether such a background is realized in the universe, 
we have to derive an effective potential for the dilaton,
and analyze its stability.
Unfortunately, 
in a pure gravitational system without quantum effects, 
there is no solution to stabilize
a dilaton.

When we discuss dynamical evolution of the universe
just after compactification,
an inflationary expansion is required in our external spacetime.
In RS setup, in which we assume the effective five-dimensional spacetime 
with a negative cosmological constant,
the bulk spacetime may be explained naturally
if we have inflation in five-dimensional external spacetime.
However, 
it have well known that de Sitter supergravity can not arise from simple
compactification of supergravity, string or M-theory\cite{hu}.
 It arises from nonstandard way. 
For example, the $R$ symmetry group corresponding to 
conserved charge is not well defined and typically non-compact in de
Sitter spacetime. Moreover the signature of kinetic term for RR 
scalar field in de Sitter spacetime is negative\cite{hu}. 
It is far from a realistic model in a supergravity theory.
In cosmology, however, we would not need to assume a spacetime 
supersymmetry. Then some quantum effects will become important in the
dynamics of the universe.

The purpose of this paper is to investigate a 
stabilization mechanism of the internal space via quantum effects
in the ten dimensional Kaluza-Klein theory and IIB supergravity model.
We assume that the vacuum ten dimensional spacetime is 
compactified into the direct product of five dimensional (anti-)de Sitter 
spacetime and compact five dimensional sphere, i.e $(A)dS_5\times S^5$.  
We consider the quantum fluctuations associated with several matter
fields in order to stabilize the scale of internal space $S^5$.
Many works suggest that quantum correction of higher dimensional matter
field might provide a physical mechanism which is capable of accounting
 for extreme smallness of the extra dimensions.

The low energy effective action in our model is obtained by the
integration of internal space, which is often called 
dimensional reduction. 
After the dimensional reduction,  the dilaton couples the matter
 fields in external spacetime. 
The quantum fluctuations around a classical solution are computed in the
form of quantum effective potential as a function of dilaton.
 Then the quantum correction of the matter field naturally
contributes to the dilaton potential.  
Since the quantum correction is dominant effect at the small scale of
 internal space, this correction at nearly Planck scale 
is expected to be very important.

This paper is organized as follows. In \S\ref{sec:stabilization}, 
we will calculate the 1-loop quantum correction and investigate 
its property for several fields in $(A)dS_5\times S^5$. 
We will apply our approach to more realistic model, i.e. type 
IIB supergravity in $dS_5\times S^5$ in \S\ref{sec:IIBSG}. 
Conclusion  follows in \S\ref{sec:conclusion}.
In Appendix, we also present the zeta functions for the case of 
$AdS_5\times S^5$ compactification.

\section{Dynamics of dilaton in $(A)dS_5\times S^5$}
\label{sec:stabilization}
First we discuss the dynamics of extra dimension in
pure gravity system.  
We consider the ten dimensional Einstein-Hilbert action with a
cosmological constant:  
%
\begin{equation}
I_{EH}=\frac{1}{2\bar{\kappa}^2} \int d^{10}X
\sqrt{-\bar{g}}(\bar{R}-2\bar{\Lambda})\,,
\label{eq;2-1}
\end{equation}
where $ \bar{\kappa} $ is a positive constant, $\bar{R}$ is the
ten dimensional Ricci scalar, and $\bar{\Lambda}$ is the cosmological 
constant. The vacuum state is assumed to be a five dimensional 
de Sitter space ($dS_5$) with a small extra  sphere ($S^5$). 
Our ansatz for the metric is the following;
%
\begin{equation}
\bar{g}_{MN}dX^MdX^N=\left(\frac{b}{b_0}\right)^{-10/3}
g_{\mu\nu}dx^{\mu}dx^{\nu}+ b^2\Omega_{ij}^{(5)}dy^idy^j\,,
         \label{eq;2-2}
\end{equation}
where $g_{\mu\nu}$ is the metric 
of a five dimensional de Sitter spacetime,
$b$ is the scale of a five dimensional sphere 
(i.e. a radius of $S^5$), 
a constant $b_0$ is an initial value of $b$, and
$\Omega^{(5)}_{ij}dy^idy^j$ is the line element of 
a unit five dimensional sphere.
 $g_{\mu\nu}$ and $b$ depend only on the 5 dimensional coordinate
\{$x^\mu\} (\mu=0, 1, 2, \cdots)$.
According to the ansatz (\ref{eq;2-2}), we  truncate
our model to a five
dimensional effective theory, in which $b$ is so-called ``dilaton''.
Substituting the metric ansatz (\ref{eq;2-2}) into the action 
(\ref{eq;2-1}), we find the five dimensional effective action is given
by   
%
\begin{equation}
 I_{g}=\int d^5x\sqrt{-g}\left[\frac{1}{2\kappa^2}R
	-\frac{1}{2}g^{\mu\nu}\partial_{\mu}\sigma\partial_{\nu}\sigma 
	- U_{0}(\sigma)\right]\,,
\label{eq;2-3}
\end{equation}
where $\kappa$ is a positive constant defined by 
$\kappa^2=\bar{\kappa}^2/(2^5b_{0}^5\pi)$ and $R$ is a Ricci scalar 
of the five dimensional metric tensor $g_{\mu\nu}$, 
the field $\sigma$ is defined by 
%
\begin{equation}
 \kappa_\sigma \sigma  =  \ln\left(\frac{b}{b_{0}}\right)\,, 
	 \hspace{1cm}
 \kappa_\sigma  =  \sqrt{\frac{2}{35}}\kappa\,,
         \label{eq;2-4}
\end{equation}
and the potential $U_{0}(\sigma)$ of the field $\sigma$ is given by
%
\begin{equation}
 U_{0}(\sigma) = 
        \frac{\bar{\Lambda}}{\kappa^2}e^{-{10\over 3} \kappa_\sigma
\sigma }
 	-\frac{10}{\kappa^2b_{0}^2}e^{-{16\over 3} \kappa_\sigma
\sigma}\,.
         \label{eq;2-5}
\end{equation}
Since  this potential does not have any local
minimum,
there is no stable compactification by $S^{5}$.
\subsection{Quantum correction in $dS_5\times S^5$}
\label{subsec:dS5}
Next we consider the quantum matter fluctuations as an origin of an energy
momentum tensor in $dS_5\times S^5$ spacetime. 
The quantum correction arising from
matter field is very important to stabilize
the scale of extra dimension. If this correction does not exist in our
model,  it is impossible of dilaton to stabilize. Then the extra dimension
 finally collapses to singularity or expands forever. 
We have so far  a lot of work on calculation of quantum correction
in curved spacetime. 
Here we adopt  the path integral  to compute the dilaton
effective potential. Any divergence appeared in 
calculation must be removed by regularization technique. This paper uses
the zeta function regularization, which was developed for performing the
path integral in curved spacetime\cite{uz1}. To calculate the quantum
correction, we consider the 1-loop quantum correction for  several matter
fields.   In the
following, we review how it leads to Gaussian functional integrals,
which can be expressed as functional determinants. In order to evaluate
the functional integrals, we introduce the generalized zeta
function which is the sum of the operator eigenvalues. We adopt this
method to determine the 1-loop effective potential in de
Sitter spacetime .
\\
\\                                                                       
           
\begin{center}
{\bf (A) Massless scalar field}
\end{center}
First, we consider the ten dimensional massless scalar field;  
%
\begin{equation}
I_{S}=-\frac{1}{2}\int d^{10}X\sqrt{-\bar{g}}\bar{g}^{MN}
 	\partial_{M}\bar{\phi}\partial_{N}\bar{\phi}\,.
         \label{eq;2-6}
\end{equation}
The ten dimensional line element is assumed to be given by Eq. 
(\ref{eq;2-2}).
To derive a five dimensional effective action,  it is
convenient to expand the field in terms of harmonics on the five
dimensional sphere: 
%
\begin{equation}
 \bar{\phi} = b_{0}^{-5/2}
	\sum_{l,m}\phi_{lm}(x^{\mu})Y^{(5)}_{lm}(y^i)\,,
         \label{eq;2-7}
\end{equation}
where $Y^{(5)}_{lm}$  are real harmonics 
on the $5$-sphere, which satisfy 
%
\begin{eqnarray}
 \frac{1}{\sqrt{\Omega^{(5)}}}\partial_{i}\left(
 	\sqrt{\Omega^{(5)}}\Omega^{(5)ij}\partial_{j}Y^{(5)}_{lm}\right)
	 &+& l(l+4)Y^{(5)}_{lm} 
 =  0\,,\\
 \int d^5x\sqrt{\Omega^{(5)}}Y^{(5)}_{lm}Y^{(5)}_{l'm'} 
 & = & \delta_{ll'}\delta_{mm'}\,.
\end{eqnarray}
$l=1,2,\cdots$; and $m$ denote a set of four 
numbers, which is required  in order for a set of all
$Y^{(5)}_{lm}$ to be a  complete set of $L^2$ functions on the $5$-sphere,
and $\phi_{lm}(x^{\mu})$ is a real function depending only on 
the five dimensional coordinate $\left\{x^{\mu}\right\}$.

By the substituting the expansion (\ref{eq;2-7}) 
into the action (\ref{eq;2-6}), 
the five dimensional effective action is given by
%
\begin{equation}
 I_{S} =  -\frac{1}{2}\sum_{l,m}\int d^5x\sqrt{-g}
 	\left[g^{\mu\nu}\partial_{\mu}\phi_{lm}
 	\partial_{\nu}\phi_{lm} 
 	+ M_{\phi}^2\;\phi_{lm}^2
 	\right]\,,
         \label{eq;2-8}
\end{equation}
where the mass $M_{\phi}^2$ of scalar field is given by 
%
\begin{equation}
M^2_{\phi}=\frac{l(l+4)}{b_0^2}e^{-{16\over 3} 
\kappa_\sigma
\sigma}\,.
\end{equation}
  
Now we compute the quantum correction of the scalar field in 1-loop
level. 
The calculation of the effective potential is carried out using path
integral method.
The fields are split into a classical part $\phi_{lm\;c}$
and quantum part $\delta{\phi_{lm}}$. The action
is then expanded in the quantum fields around arbitrary classical
background field. We expand the fields to second order to
calculate all 1-loop diagrams with any number of lines of external
fields.
 
In the path integral approach to quantum field theory, the amplitude 
is given by an expression
%
\begin{equation}
Z=\int{\cal D}[\phi_{lm}]\exp{\left(iI_t[\sigma, \phi_{lm}]\right)},
       \label{eq;subquantum-1}
\end{equation}
where ${\cal D}[\phi_{lm}]$ is a measure on the functional 
 space of scalar fields, and $I_{t}[\sigma, \phi_{lm}]$ is the total action. 
  
The action is expanded in a neighborhood of these classical
 background fields as follows:
%
\begin{equation}
I_{t}[\sigma,\phi_{lm}]=I[\sigma, \phi_{lm\;c}]+I_q[\sigma, \delta{\phi_{lm}}]
                   +O\left((\delta\phi)^3\right),
       \label{eq;subquantum-2}       
\end{equation}
where $\phi_{lm}=\phi_{lm\;c}+\delta\phi_{lm}$. The action 
$I_{t}[\sigma,\phi_{lm}]$ is  
quadratic in $\delta\phi_{lm}$. The linear terms of $\delta\phi_{lm}$
has disappeared due to the classical equations of motion.  We neglect 
all higher order terms than quadratic one in  the 1-loop
approximation. Then, the   expression becomes  
%
\begin{equation}
\ln Z=iI[\sigma, \phi_{lm\:c}]
   +\ln\left\{\int{\cal D}[\delta\phi_{lm}]
     \exp{\left(iI_q[\sigma, \delta\phi_{lm}]\right)}\right\}.
          \label{eq;3-2-2}
\end{equation}
We note that the integral is ill-defined because the operators in Eq.
(\ref{eq;3-2-2}) are unbounded from below  in the $dS_5$ spacetime
 with Lorentz
signature. We have to perform a Wick rotation in order to redefine it
well  and rewrite 
it in the Euclidean  form.
We then obtain  the expression
%
\begin{eqnarray}
\ln Z&=&-I_E[\sigma, \phi_{lm\;c}]\nonumber\\
      &+&\ln\left\{\int{\cal D}[\delta\phi_{lm}]
       \exp{\left(-I_{qE}[\sigma, \delta\phi_{lm}]\right)
       }\right\},
          \label{eq;3-2-3}
\end{eqnarray}
where $I_E$ is the Euclidean action expressed by  
%
\begin{equation}
I_E[\sigma, \phi_{lm}]=\frac{1}{2}\sum_{l,\;m}\int d^5 x\sqrt{-g}
             \left(\partial_{\mu}\phi_{lm}
             \partial^{\mu}\phi_{lm}+M^2_{\phi}\phi^2_{lm}\right).
          \label{eq;3-2-4}
\end{equation}
Using the assumption $\phi_{lm}=\phi_{lm\;c}+\delta\phi_{lm}$,
 we can integrate the kinetic term in the action by parts, resulting in 
%
\begin{equation}
I_E[\sigma, \delta\phi_{lm}]=\frac{1}{2}\int d^5 x\sqrt{-g}
             ~\delta\phi_{lm}\left\{-\nabla^2_{(5)}
              +M^2_{\phi}\right\}\delta\phi_{lm},
           \label{eq;3-2-5}
\end{equation}
where $\nabla^2_{(5)}$ denotes the Laplacian in the five dimensional
 de Sitter spacetime.  
 
The effective potential due to quantum correction ($V_{qc}$)
 is defined by the relation  
%
\begin{eqnarray}
\exp\left(-\int d^5x V_{qc}\right)
   &=&\int {\cal D}\left[\phi_{lm}\right]
    \exp\left(-I_{qE}[\sigma, \delta\phi_{lm}]\right)\nonumber\\
   & &\hspace{-1cm}
     =\left\{\det\mu^{-2}\left(\nabla^2_{(5)}-M^2_{\phi}\right)
     \right\}^{-\frac{1}{2}},
  \label{eq;3-2-6}
\end{eqnarray}
where  $\mu$ is a 
normalization constant with  mass dimension.
To compute the effective potential to 1-loop, we define it by 
%
\begin{equation}
Z=\exp\left(-\int d^5x V_{eff}\right)
   =\exp\left(-\Omega_{vol}V_{eff}\right),
  \label{eq;3-2-7}
\end{equation}
where $\Omega_{vol}$ is the volume of five dimensional de Sitter
 spacetime.

Using the Eqs.(\ref{eq;3-2-3}),(\ref{eq;3-2-6}),(\ref{eq;3-2-7}), 
we find that 1-loop effective potential is 
%
\begin{eqnarray}
V_{eff}(\sigma)
&=&U_0(\sigma)+V_{qc}(\sigma)
\nonumber \\
& &\hspace{-1.4cm}=U_0(\sigma)+\frac{1}{2\Omega_{vol}}
             \ln\det\left\{\mu^{-2}\left(\nabla^2_{(5)}
             -M^2_{\phi}\right)\right\}.
  \label{eq;3-2-8}
\end{eqnarray}

 We shall evaluate the functional determinant  on a background manifold 
in $dS_5$.
 We apply the standard technique of zeta function regularization.
The functional determinants in terms of the generalized 
zeta function is the sum of operator eigenvalue 
%
\begin{eqnarray}
\zeta_{\phi}(s)&\equiv&\sum^{\infty}_{l=0}\sum^{\infty}_{l'=0}
           \frac{2(l+2)(l+3)!}{4!\;l!}\:d_{\phi}(l')
           \nonumber\\     
    & &\times \left\{\frac{\lambda_{\phi}(l')}{a^2}
          +\frac{l(l+1)}{b_0^2}e^{-{10\over 3} \kappa_\sigma\sigma }
          \right\}^{-s}, 
  \label{eq;3-3-1}
\end{eqnarray}
where $a$ is the scale of $dS_5$, and $\lambda_{\phi}(l')$
is the eigenvalue of scalar field $\phi$ on
 $dS_5$, and $d_{\phi}(l')$ is its degeneracy,  respectively.
This expansion is well defined and converge for $Re(s)>5$.
Using this function, the effective potential (\ref{eq;3-2-8}) is written
by 
%
\begin{equation}
V_{eff}(\sigma)=U_0(\sigma)-\frac{1}{2\Omega_{vol}}
             \left\{{\zeta_{\phi}}'(0)
             +\zeta_{\phi}(0)\ln({\mu}^2a^2)\right\},      
  \label{eq;3-3-2}
\end{equation}
where in order to get the second term, we have used the relation
%
\begin{equation}
\det\left(\mu M\right)=\mu^{\zeta(0)}\det M.
      \label{eq;zr}
\end{equation}
Our task is now to calculate $\zeta_{\phi}(s)$ and to analytically
 continue to
 $s=0$ to evaluate $\zeta_\phi(0)$ and $\zeta_\phi^{~\prime}(0)$.
Giving eigenvalue of operator 
in de Sitter spacetime,
the zeta function $\zeta_{\phi}$ is 
 evaluated.
The $dS_5$ spacetime is a five dimensional hyperboloid with a  constant
 curvature and has a unique Euclidean section
$S^5$ with a  radius $a$. 
The degeneracy $d_{\phi}(l)$ and 
the eigenvalue $\lambda_{\phi}(l)$ of 
massless scalar field in $dS_5$ spacetime are well known because this
spacetime  equal to $S^5$ in the Euclidean section\cite{ru}.
These are given by
%
\begin{eqnarray}
d_{\phi}(l')=l'(l'+1),\hspace{0.5cm}
\lambda_{\phi}(l')=\frac{2(l'+2)(l'+3)!}{4!(l'!)}.
\end{eqnarray}
 
If the condition of $a\gg b$ is satisfied, we  easily calculate
 the value
of
$\zeta_{\phi}(0)$ and $\zeta_{\phi}'(0)$.
Then the effective potential is given by 
%
\begin{eqnarray}
V_{eff}&=&\frac{\bar{\Lambda}}{\kappa^2}
 e^{-{10\over 3}\kappa_\sigma\sigma}
        -\frac{10}{\kappa^2 b_0^2}e^{-{16\over 3}\kappa_\sigma\sigma}
      +\frac{1}{b_0^5}e^{-{40\over 3}\kappa_\sigma\sigma}
  \nonumber \\
& \times &
\left[
 -\ln(\mu^2a^2)
        +C_{\phi}\ln
\left\{
\left(
\frac{a}{b_0}
\right)^2
        e^{-{16\over 3}\kappa_\sigma\sigma}
\right\}
\right] ,  
  \label{eq;3-3-3}
\end{eqnarray}
where $C_{\phi}$ is given by 
%
\begin{eqnarray}
C_{\phi}&=&\frac{15}{16\pi^2}\zeta_{\phi}'(0)
           \:e^{{40\over 3}\kappa_\sigma\sigma}
        =\frac{25}{8192\pi}.
  \label{eq;3-constant}
\end{eqnarray}

\begin{center}
{\bf (B) $U(1)$ gauge field }
\end{center}
Next we compute the quantum correction of $U(1)$ gauge field $A_M$
described the action;
%
\begin{equation}
S_{U(1)}=\int d^{10}X\sqrt{-\bar{g}}F_{MN}F^{MN},
        \label{eq;elac}
\end{equation}
where $F_{MN}=\nabla_M A_N-\nabla_N A_M$. 
In order to perform the dimensional reduction for 
 the $U(1)$ field action in $dS_5\otimes S^5$
 spacetime, it is convenient to
 expand it by the vector harmonics on the $S^5$ as:
%
\begin{eqnarray}
\bar{A}_MdX^M&=&b_0^{-\frac{5}{2}}\sum_{l,\:m}\left[
              A^{(5)}_{\mu\;lm}Y^{(5)}_{lm}
       dx^{\mu}+\left\{A^{(5)}_{(T)\;lm}\left(V^{(5)}_{(T)\;lm}\right)_i
      \right.\right. \nonumber \\
&&\left.\left.+A^{(5)}_{(L)\;lm}\left(V^{(5)}_{(L)\;lm}\right)_i\right\}dy^i
              \right],
\label{eq;expand}
\end{eqnarray}
where $A^{(5)}_{\mu\;lm}$, $A^{(5)}_{(T)\;lm}$ and $A^{(5)}_{(L)\;lm}$
depend only on the five dimensional
coordinate ${x^{\mu}}$. 
$Y^{(5)}_{lm}$ and $V^{(5)}_{(T)\;lm}$, 
$V^{(5)}_{(L)\;lm}$ are the scalar
harmonics, transverse vector harmonics, and longitudinal vector
harmonics respectively. 
As $A^{(5)}_{(L)\;lm}$ represents gauge degrees of freedom,  we eliminate
them after the gauge fixing 
(See the Appendix in Ref.\cite{uz3} for definition 
and properties of there harmonics).
By substituting the expansion (\ref{eq;expand})
 into the action (\ref{eq;elac}), we find the five dimensional effective
 action. 

As this effective action still has dilaton coupling for
 vector and  scalar modes,
in order to evaluate the eigenvalues 
in the path integral, 
we integrate it by part and then rewrite 
the integrand to the proper form. We 
divide the field 
$A_{\mu}$ to the transverse and longitudinal parts as,
%
\begin{equation}
A^{(5)}_{\mu}dx^{\mu}=\left(A^{(5)}_{(T)\:\mu}
              +A^{(5)}_{(L)\;\mu}\right)dx^{\mu}.
     \label{eq;3c-5}
\end{equation}
For the quantization of $U(1)$ gauge field $A^{(5)}_{\mu}$, we choose a
Lorentz gauge. The action for $U(1)$ gauge field
$A^{(5)}_{\mu}$ is finally given by 
%
\begin{eqnarray}
S_{U(1)} & = & S_{(V)}+S_{(T)}+\delta S,\nonumber\\
S_{(V)} & = &-\frac{1}{2}\int d^5x\:\sqrt{-g}\:
A^{(5)\;\mu}_{(T)\;lm}
\nonumber \\
&&
\left\{\frac{1}{2}
\left(g_{\mu\nu}\nabla^2_{(5)}-\nabla_{\mu}\nabla_{\nu}\right)
e^{-{10\over 3}\kappa_\sigma\sigma}
\right.
\nonumber \\
&& \left.
-e^{-{10\over 3}\kappa_\sigma\sigma}
\bigtriangleup_{\mu\nu}
 +g_{\mu\nu}M^2_{(V)}\right\}A^{(5)\;\nu}_{(T)\;lm},\nonumber\\
S_{(T)} & = &-\frac{1}{2}\int d^5x\:\sqrt{-g}\:
A^{(5)}_{(T)\;lm}\left\{
\frac{1}{2}\nabla^2_{(5)} e^{-2\kappa_\sigma\sigma}
\right.
\nonumber \\
&&
\left.
   -e^{-2\kappa_\sigma\sigma}\nabla^2+M^2_{(T)}
\right\}A^{(5)}_{(T)\;lm},   \nonumber\\
\delta S & = & -\frac{1}{2}\int d^5x\:\sqrt{-g}\:A^{(5)\;\mu}_{(L)\;lm}
\nabla_{\mu}\nabla_{\nu} A^{(5)\;\nu}_{(L)\;lm},
\label{eq;vaction}
\end{eqnarray}
where $\delta S$ is the gauge fixing action and $\alpha$ is 
positive constant and 
$\bigtriangleup_{\mu\nu}=g_{\mu\nu}\nabla^2_{(5)}+R_{\mu\nu}$. 
$M^2_{(V)}$ and $M^2_{(T)}$ are mass of the five dimensional vector field
 $A^{(5)\;\mu}_{(T)\;lm}$ and that of the five
 dimensional scalar field $A^{(5)}_{(T)\;lm}$, respectively, which  are
given by   
%
\begin{eqnarray}
M^2_{(V)}&=&e^{-2\kappa_\sigma\sigma}
             \left\{\frac{l(l+4)}{b_0^2}\right\},\\
M^2_{(T)}&=&e^{-{22\over 3}\kappa_\sigma\sigma}
             \left\{\frac{l(l+4)+3}{b_0^4}\right\}.
\end{eqnarray}
To calculate the 1-loop quantum correction by the $U(1)$ gauge field, 
we follow the procedure discussed in \cite{uz1}. 
Finally, we obtain the effective potential due to  
$A^{(5)\;\mu}_{(T)\;lm}$  and $A^{(5)}_{(T)\;lm}$ as follows:
\begin{eqnarray}
&&V_{U(1)\;eff} = \frac{\Lambda}{\kappa^2}
e^{-{10\over 3}\kappa_\sigma\sigma}
        -\frac{10}{\kappa^2 b_0^2}
e^{-{16\over 3}\kappa_\sigma\sigma}
    +  \frac{1}{b_0^5}e^{-{40\over 3}\kappa_\sigma\sigma}  \nonumber \\
 &&~~\times
        \left[-\ln(\mu^2a^2)+C_V\ln\left\{\left(\frac{a}{b_0}\right)^2
e^{-{16\over 3}\kappa_\sigma\sigma}
\right\}\right]
\,,
\end{eqnarray}
where $C_V$ is given by 
%
\begin{eqnarray}
C_{V}=\frac{15}{16\pi^2}\zeta_{V}'(0)
        \:e^{{40\over 3}\kappa_\sigma\sigma}
       =\frac{1199}{49152\pi},
\end{eqnarray}
where
%
\begin{equation}
\zeta_{V}(s)
        =\sum^{\infty}_{l=0}\sum^{\infty}_{l'=1}D_{V}(l)d_{V}(l')
        \left[e^{-{10\over 3}\kappa_\sigma\sigma}
            \frac{\lambda_{V}(l')}{a^2}
         +e^{-{2}\kappa_\sigma\sigma}\Lambda_{\chi}^2\right],
       \label{eq;3B-14}
\end{equation}
where degeneracies $D_{V}(l)$, $d_{V}(l')$ and eigenvalues
$\lambda_{V}(l')$, $\Lambda_{V}(l)$ are  given by 
%
\begin{eqnarray}
D_{V}(l)&=&\frac{(l+1)(l+2)(l+3)}{12},\nonumber\\
d_{V}(l')&=&\frac{l}{3}(l+2)^2(l+4), \nonumber\\ 
\Lambda^2_{\chi}(l)&=&\frac{l(l+4)}{b_0^2},\nonumber\\
\lambda^2_{\chi}(l')&=&\left\{l(l+4)-3\right\}. 
\end{eqnarray}

\begin{center}
{\bf (C) Dirac spinor field }
\end{center}
We then calculate the quantum correction associated with massless Dirac
spinor field on $dS_5\times S^5$. 
The action is given by 
%
\begin{equation}
S_{\psi}=i\int d^{10}X \sqrt{-g}\left(\gamma^M\nabla_M \psi\right). 
\end{equation}
The spinor representation of O$(1,\;5+5)$ is a  direct product of the
spinor representation of O$(1,\;4)$ and O$(5)$.
The ten dimensional gamma matrix $\bar{\gamma}$ is given by 
%
\begin{eqnarray}
&&
\bar{\gamma}^{\mu}=\gamma^{\mu}\otimes {\bf 1},\hspace{.5cm}
\bar{\gamma}^i=\gamma^5\otimes\gamma^i,\hspace{.5cm}
\left(\gamma^5\right)^2=1,
\nonumber \\
&&
\left\{\bar{\gamma}^M,\;\bar{\gamma}^N\right\}=2g^{MN},
\end{eqnarray}
where the $\gamma^{\mu}\:(\mu=0,\:1,\:2,\:3,\;4)$ are Dirac matrices in 
$dS_5$ while the $\gamma^i\:(i=5,\:6,\:\cdots,\:9)$ are those in
$S^5$\cite{can}. The Dirac spinor field $\psi$ is expanded as spinor
harmonics
 analogous to scalar field;
%
\begin{equation}
 \bar{\psi}(x^{\mu},\;y^i) = b_{0}^{-5/2}
	\sum_{l,m}\psi_{lm}(x^{\mu})Y^{(5)}_{\psi\;lm}(y^i),
         \label{eq;fhar}
\end{equation}
where $\psi_{lm}(x^{\mu})$ is the Dirac spinor field in the five 
dimensional spacetime.
$Y^{(5)}_{\psi\;lm}$ 
are real spinor harmonics on the $S^5$ satisfying 
%
\begin{eqnarray}
i\:\bar{\gamma}^i\nabla_i\;Y^{(5)}_{\psi\;lm}
 & = &\Lambda_{\psi}Y^{(5)}_{\psi\;lm} ,\\
 \int d^5y\sqrt{\Omega^{(5)}}\:Y^{(5)}_{\psi\;lm}\:Y^{(5)}_{\psi\;l'm'} 
 & = & \delta_{ll'}\delta_{mm'},
\end{eqnarray}
and $\psi_{lm}(x^{\mu})$ is a real function depending only on 
the five dimensional coordinates $x^{\mu}$.  $l=1,2,\cdots$, and $m$
denote a set of six   numbers which is required  in order for
$Y^{(5)}_{\psi\;lm}$ to be a complete set of $L^2$ functions on  the
$S^5$)  Here $\bar{\gamma}^i\nabla_i$ is the Dirac operator on the unit
seven  sphere 
$S^5$ and $\Lambda_{\psi}(l)$ denotes a eigenvalue for the Dirac spinor
field $\psi$.
Using the relation for the ten dimensional Dirac operator 
$\bar{\gamma}^M\nabla_M=\bar{\gamma}^{\mu}\nabla_{\mu}\otimes {\bf 1}
                       +\gamma^5\otimes\bar{\gamma}^i\nabla_i$, 
 we obtain the five dimensional effective action
%
\begin{eqnarray} 
&&
S\left(\psi_{lm},\;\bar{\psi}_{lm}\right)
=
  \sum_{l,m}\int d^5x \:\sqrt{-g}
\nonumber \\
&&
~~~~~\times\bar{\psi}_{lm}
  \left(i\gamma^{\mu}\nabla_{\mu}+\Lambda_{\psi}\:\gamma^5\right)
  \psi_{lm}.
\end{eqnarray}
The partition function $Z$ for massless Dirac spinor field on $dS_5$ is 
%
\begin{equation}
Z=\int {\cal D}\left[\psi_{lm}\right]{\cal D}\left[\bar{\psi}_{lm}\right]
 \exp\left\{-iS\left(\psi_{lm},\;\bar{\psi}_{lm}\right)\right\},
\end{equation}
where ${\cal D}\left[\psi_{lm}\right]$ and
${\cal D}\left[\bar{\psi}_{lm}\right]$ are the functional measure of
the spinor field $\psi_{lm}$ and its Dirac adjoint field 
$\bar{\psi}_{lm}$, respectively.
Using the definition of a Gaussian functional for anti-commuting fields,
we obtain the partition function as
%
\begin{eqnarray}
\ln Z&=&\ln\det\left\{\mu^{-1}
  \left(i\gamma^{\mu}\nabla_{\mu}
  +\Lambda_{\psi}(l)\gamma^5\right)\right\}\nonumber\\
 &=&\frac{1}{2}\ln\det\left[\mu^{-2}
   \left\{-\left(\gamma^{\mu}\nabla_{\mu}\right)^2
  +\left(\Lambda_{\psi}(l)\right)^2\right\}\right],
\nonumber \\
~
\end{eqnarray}
where five dimensional Dirac operator 
$\left(\gamma^{\mu}\nabla_{\mu}\right)^2$ is given by \cite{sc,li}
%
\begin{equation}
\left(\gamma^{\mu}\nabla_{\mu}\right)^2=\nabla^2_{(5)}+\frac{1}{4}R_{(5)}.
\end{equation}
The effective potential is then rewritten as
%
\begin{equation}
V_{eff}(\sigma)=U_0(\sigma)-\frac{1}{4\Omega_{vol}}
\left\{
\zeta_f'(0) +\zeta_f(0)\ln(\mu^2a^2)
\right\},
\end{equation}
where $\zeta_f(0)$
 is the generalized zeta function for the Dirac 
spinor field:
%
\begin{equation}
\zeta_{\psi}(s)
        =\sum^{\infty}_{l=2}\sum^{\infty}_{l'=0}D_{\psi}(l)d_{\psi}(l')
        \left[\frac{\lambda_{\psi}(l')}{a^2}
         +\frac{\Lambda_{\psi}^2}{b_0^2}\right].
       \label{eq;3C-14}
\end{equation}
Here degeneracies $D_{\psi}(l)$, $d_{\psi}(l')$ and eigenvalues
$\lambda_{\psi}(l')$, $\Lambda_{\psi}(l)$ are  given by 
%
\begin{eqnarray}
D_{\psi}(l)&=&\frac{(l+4)(l+3)(l+2)(l+1)}{4!},\nonumber\\
d_{\psi}(l')&=&\frac{(l'+4)(l'+3)(l'+2)(l'+1)}{4!}, \nonumber\\ 
\Lambda^2_{\psi}(l)
    &=&e^{-{16\over 3}\kappa_\sigma\sigma}
  \left(l+\frac{5}{2}\right)^2,\nonumber\\
\lambda^2_{\psi}(l')&=&\left(l'+\frac{5}{2}\right)^2-5. 
\end{eqnarray}
Following the  method given in \cite{uz1}, we find
the effective potential to 1-loop order is 
%
\begin{eqnarray}
V_{eff}&=&
\frac{\Lambda}{\kappa^2}
e^{-{10\over 3}\kappa_\sigma\sigma}
        -\frac{10}{\kappa^2 b_0^2}
e^{-{16\over 3}\kappa_\sigma\sigma}
  +  \frac{C_{f}}{b_0^5}e^{-{40\over 3}\kappa_\sigma\sigma}  \nonumber \\
\,,
\end{eqnarray}
where $C_f$ is given by 
%
\begin{eqnarray}
C_{f}&=&\frac{15}{16\pi^2}\zeta_{f}'(0)
        \:e^{{40\over 3}\kappa_\sigma\sigma}\nonumber\\
    &=&-3.335245\times 10^{-6}\,.
\end{eqnarray}
Note that the logarithmic term does not appear because $\zeta_f(0)$
vanishes. The same problem arises in Minkowski spacetime\cite{gl}. 

\begin{center}
{\bf (D) gravitational field  (scalar mode)}
\end{center}

Finally, we investigate the quantum correction by
 gravitational field. 
In our model, it is assumed that the distance scale of the
 extra dimension is by a few order magnitude larger than Planck scale.  
We then apply the method of the conventional loop expansion approach to
the
 quantization of the gravitational field theory\footnote{At the Planck
 length, however, the loop expansion will break down because the
 effective self-coupling will becomes the order unity.}.   

 We consider a gravitational perturbation $h_{MN}$ around a
background metric $\bar{g}_{MN}^{(0)}$, which we shall specify later; 
%
\begin{equation}
\bar{g}_{MN} =\bar{g}_{MN}^{(0)}+h_{MN}.
\label{eqn;2}
\end{equation}
Substituting Eq.(\ref{eqn;2}) into Eq.(\ref{eq;2-1}), we obtain the
perturbed Einstein-Hilbert action as follows. 

%
\begin{eqnarray}
 I_{EH} = 
 	\int d^{10}x\sqrt{-\bar{g}^{(0)}} \left[{\cal L}_{(0)}+{\cal L}_{(2)}\right],
		\label{eqn:perturbedEH}
\end{eqnarray}
where
\begin{eqnarray}
{\cal L}_{(0)} & =  & \frac{1}{2\bar{\kappa}^2} 
	\left[ \bar{R}^{(0)} -2\bar{\Lambda} \right]
\label{eqn:perturbedEH0}\\
{\cal L}_{(2)}&=&\frac{1}{2\bar{\kappa}^2} \left[
\frac{1}{8}\left( h^2-2h^{MN}h_{MN}\right)\bar{R}^{(0)} 
	\right.
\nonumber \\
&&
\left.
+\frac{1}{2}\left(2h^{MM'}{h_{M'}}^N-hh^{MN}\right)
	\bar{R}^{(0)}_{MN}  \right.\nonumber  \\ 
 &&  + \left. \frac{1}{4} \left\{ {h^{MN}}_{;M'}
	\left(2{h^{M'}}_{M;N}- {h_{MN}}^{;M'}\right) 
 \right.\right.
\nonumber \\
&& \left.	+ h_{;M} \left(h^{;M}-2{h^{MN}}_{;N}\right) 
	\right\}  
\nonumber \\
&&	\left.
-\bar{\Lambda}\left(\frac{1}{4}h^2-\frac{1}{2}h^{MN}h_{MN} 
	\right)  +O(h^3)\right].
	\label{eqn:perturbedEH2}
\end{eqnarray}
where ``$;$'' denotes the covariant derivative with respect to 
$g^{(0)}_{MN}$, and $\bar{R}^{(0)}_{MN}$ and $\bar{R}^{(0)}$ are 
the Ricci tensor and scalar of the back ground metric
$\bar{g}^{(0)}_{MN}$.
As for the background geometry, we compactify it on a five dimensional
sphere $S^5$.

After gauge-fixing and
redefining $g_{\mu\nu}$ and $b$, the perturbation $h_{MN}$ 	is expanded 
as\cite{uz3}:  
%

\begin{eqnarray}
 &~&h_{MN}dX^MdX^N 
\nonumber \\
&~&= \sum_{l,\;m}
	\left[ h_{\mu\nu}^{lm}Y_{lm}dx^{\mu}dx^{\nu}
	+ 2h_{(T)\mu}^{lm}(V_{(T)lm})_idx^{\mu}dy^i
\nonumber \right.
\\
&~&\left.
	+ \{ h_{(T)}^{lm}(T_{(T)lm})_{ij}
	+ h_{(Y)}^{lm}(T_{(Y)lm})_{ij}\}dy^idy^j\right],
	\label{eqn:gaugefixedh}
\end{eqnarray}
where $Y_{lm}$, $V_{(T)lm}$, $T_{(T)lm}$, and 
$T_{(Y)lm}$ are the scalar harmonics, the vector harmonics, and two tensor
harmonics, respectively. The  coefficients $h_{\mu\nu}^{lm}$,
$h_{(T)\mu}^{lm}$,  
$h_{(T)}^{lm}$, and $h_{(Y)}^{lm}$
depend only on the five dimensional coordinates $x^{\mu}$, while the
harmonics depend only on the coordinates $y^i$ on $S^5$.
The summations are taken over $l\ge 1$ for the scalar and vector 
harmonics, and over $l\ge 2$ for the tensor harmonics. 

Herewith we only calculate the scalar mode $h_{(T)}^{lm}$,
just for a technical reason.
Substituting the above harmonic expansion into the Einstein-Hilbert action
(\ref{eqn:perturbedEH}), we then obtain  the following action.
%
\begin{equation}
I = I_{g} + I_{\chi},
\end{equation}
where $I_{g}$ is five dimensional Einstein-Hilbert action(\ref{eq;2-3})
and 
%
\begin{eqnarray}
I_{\chi}  &= & -\frac{1}{2}\int d^5x\sqrt{-\bar{g}^{(0)}}\left[
e^{-4\kappa_\sigma\sigma}
g^{\mu\nu}\partial_{\mu}\chi^{lm}\partial_{\nu}\chi^{lm}
\nonumber \right.\\
&&\left.
+M^2_{\chi\;lm}\chi^{lm}\chi^{lm}\right].
       \label{eq;3D-12}
\end{eqnarray}
The scalar mode field $\chi_{lm}$ and its mass $M^2_{\chi}$
are defined as follows: 
%
\begin{eqnarray}
 M^{2}_{(\chi)\;lm} & = &
 e^{-{28\over 3}\kappa_\sigma\sigma}
 \{l(l+4)+12\}
 b_0^{-2}\nonumber\\
 & &+
 e^{-4\kappa_\sigma\sigma}
\left[
 \frac{22}{3}\kappa_\sigma\nabla^2_{(5)}
 \sigma
 -\frac{28}{3}\kappa_\sigma^2\left(\nabla_{(5)}\;
 \sigma\right)^2 
\nonumber\right.\\
&&\left.
 +
\left\{
 R^{(0)}-2
 e^{-{10\over 3}\kappa_\sigma\sigma}
 \bar{\Lambda}
\right\}
\right]
,\\
\chi^{lm} & \equiv &
\frac{1}{8\sqrt{2\pi}\kappa b_0^{2}}
	\;h^{lm}_{(T)} \qquad (l\ge 2).
       \label{eq;3D-11}
\end{eqnarray}
In order to analyze quantum correction, we rewrite the above  expression 
because of the existence of dilaton coupling  to kinetic term of $\chi$.
After some calculation, we obtain 

%
\begin{eqnarray}
I_{\chi}&=&-\frac{1}{2}\int d^5x \sqrt{-\bar{g}^{(0)}}\chi_{lm}\left[
     \frac{1}{2}\nabla^2_{(5)} 
e^{-4\kappa_\sigma\sigma}   
\nonumber \right.\\
&& \left.
-e^{-4\kappa_\sigma\sigma}
\nabla^2_{(5)}
    +M^2_{\chi}\right]\chi_{lm}.
       \label{eq;3D-13}
\end{eqnarray}

In order to compute the 1-loop quantum correction, we introduce the zeta
function as
%
\begin{equation}
\zeta_{\chi}(s)
        =\sum^{\infty}_{l=2}\sum^{\infty}_{l'=0}D_{\chi}(l)d_{\chi}(l')
        \left[\frac{\lambda_{\chi}(l')}{a^2}
    e^{-4\kappa_\sigma\sigma}     
         +
e^{-{28\over 3}\kappa_\sigma\sigma}
\Lambda_{\chi}^2\right],
       \label{eq;3D-14}
\end{equation}
where degeneracies $D_{\chi}(l)$, $d_{\chi}(l')$ and eigenvalues
$\lambda_{\chi}(l')$, $\Lambda_{\chi}(l)$ are  given by 
%
\begin{eqnarray}
D_{\chi}(l)&=&\frac{3(l-1)(l+2)^2(l+5)}{4},\nonumber\\
d_{\chi}(l')&=&\frac{1}{12}(l+1)(l+2)^2(l+3), \nonumber\\ 
\Lambda^2_{\chi}(l)&=&\frac{l(l+4)+12}{b_0^2}
         +e^{{16\over 3}\kappa_\sigma\sigma}
          \left\{\frac{20}{a^2}
         -2\Lambda e^{-{10\over 3}\kappa_\sigma\sigma}
\right\},\nonumber\\
\lambda^2_{\chi}(l')&=&\frac{1}{2}l(l+4). 
\end{eqnarray}

Using the results of Appendix \ref{sec:scalar} in \cite{uz1},
 we can evaluate 
$\zeta(s)$ and $\zeta'(s)$ for  $\chi$ at $s=0$. 
Under the assumption of $a\gg b$, 
we get the following expression; 
%
\begin{eqnarray}
&&
V_{eff}(b)=\frac{\bar{\Lambda}}{\kappa^2}
          e^{-{10\over 3}\kappa_\sigma\sigma}
 	   -\frac{21}{\kappa^2b_{0}^2}
        e^{-{16\over 3}\kappa_\sigma\sigma}
    +\frac{1}{b_0^5} e^{-{40\over 3}\kappa_\sigma\sigma}  \nonumber \\
&&~~~\times
            \left[-\ln\left(\mu^2a^2\right)
           +C_{\chi}\ln\left\{
           \left(\frac{a}{b_0}\right)^2
e^{-{16\over 3}\kappa_\sigma\sigma}
\right\}\right],
       \label{eq;3D-17}
\end{eqnarray}
where $C_{\chi}$ is given by 
%
\begin{eqnarray}
C_{\chi}&=&\frac{15}{16\pi^2}\zeta_{\chi}'(0)
        \:e^{{40\over 3}\kappa_\sigma\sigma}
  =\frac{111}{16\pi}.
\end{eqnarray}
This result indicates that one-loop graviton contributions are roughly
three orders of magnitude bigger than scalar contribution. These are
well-known result in Minkowski spacetime\cite{ch}.  
\subsection{Quantum correction in $AdS_5\times S^5$}
\label{subsec:AdS5}
As we will discuss the detail in next subsection,
the  minimum of the effective potential turns out to be 
negative in most plausible cases. 
Therefore, in order to show a consistency
of our results, 
we have to calculate the quantum correction 
in $AdS_5\times S^5$ back ground and derive the effective potential.

Since the procedure to get the effective potential in $AdS_5\times S^5$
is almost the same as that in $dS_5\times S^5$
except for the zeta functions, we do not repeat it here.
Instead, showing how to derive the zeta functions in
 $AdS_5\times S^5$ in Appendix,
we just summarize our results as follows:

The effective potentials for bosonic fields (scalar field $\bar{\phi}$,
U(1) gauge field $A_M$, and scalar mode of gravitational field $h^{lm}_{(T)}$)
is given by 
%
\begin{eqnarray}
&&V_{eff}(\sigma)=\frac{\bar{\Lambda}}{\kappa^2}
          e^{-{10\over 3}\kappa_\sigma\sigma}
 	   -\frac{10}{\kappa^2b_{0}^2}
        e^{-{16\over 3}\kappa_\sigma\sigma}   
           +\frac{1}{b_0^5}
e^{-{40\over 3}\kappa_\sigma\sigma}
          \nonumber \\
&&~~ \times\left[-\ln\left(\mu^2a^2\right)
           +D_{c}\ln\left\{
           \left(\frac{a}{b_0}\right)^2
          e^{-{16\over 3}\kappa_\sigma\sigma}
\right\}\right],
\label{QC_AdS}
\end{eqnarray}
where $D_{c}$ is given by 
%
\begin{eqnarray}
D_c
&& = \frac{\pi}{4320} ~~~{\rm for~a~scalar~field}\nonumber \\
&& = \frac{47\pi}{55296} ~~~{\rm for~a~gauge~field}\nonumber \\
&& = \frac{11\pi}{10}~~~{\rm for~a~gravitational~field}.
\end{eqnarray}
For a Dirac spinor field, the effective potential is given by 
%
\begin{eqnarray}
V_{eff}(\sigma)=\frac{\bar{\Lambda}}{\kappa^2}
          e^{-{10\over 3}\kappa_\sigma\sigma}
 	   -\frac{10}{\kappa^2b_{0}^2}
        e^{-{16\over 3}\kappa_\sigma\sigma}   
           +\frac{D_f}{b_0^5}
e^{-{40\over 3}\kappa_\sigma\sigma},\nonumber\\
\label{QC2_AdS}
\end{eqnarray}
where $D_f=-7.29353\times10^{-6}$\,.
%
 
\subsection{Dynamics of dilaton and stabilization of $S^5$}
 \label{subsec;stability}
Once we know the effective potential for the dilaton, 
it is easy to analyze a stability of extra dimensions\cite{maeda}.
To stabilize
the extra dimensions ($S^5$),  the dilaton  potential has to have  a
minimum or at least a local minimum.  From \S\ref{subsec:dS5}, 
the effective potential in $dS_5\times S^5$ is given by 
%
\begin{eqnarray}
V_{eff}(\sigma)&=&\frac{\bar{\Lambda}}{\kappa^2}
          e^{-{10\over 3}\kappa_\sigma\sigma}
 	   -\frac{10}{\kappa^2b_{0}^2}
        e^{-{16\over 3}\kappa_\sigma\sigma}\nonumber\\   
       & &+\frac{{\cal C}}{b_0^5}
 e^{-{40\over 3}\kappa_\sigma\sigma}
 +\frac{C_F}{b_0^5}
 e^{-{40\over 3}\kappa_\sigma\sigma}\,,
\label{effective_pot_dS}
\end{eqnarray}
where ${\cal C}$ and $C_F$ are given by 
%
\begin{eqnarray}
{\cal C}_B&=&\left[-N\ln\left
(\mu^2a^2
\right) +C\ln\left\{
           \left(\frac{a}{b_0}e^{-{16\over
3}\kappa_\sigma\sigma}\right)^2\right\}
\right]\,,
\end{eqnarray}
with $N=N_\phi+N_V+N_f+1$, $C=N_\phi C_\phi+N_V C_V+C_\chi$ 
and $C_F=N_fC_f$.
$N_\phi, N_V$, and $N_f$ are numbers of scalar, vector and fermion field.

In $AdS_5\times S^5$ background, the effective potential is
given by Eq.(\ref{effective_pot_dS}) with replacing the constants 
$C_\phi, C_V, C_\chi$ and 
$C_f$ with $D_\phi, D_V, D_\chi$ and  
$D_f$, respectively.
The  stable point $\sigma= \sigma_s$ must satisfy the following
conditions,
%
\begin{equation}
\frac{\partial V_{eff}}{\partial \sigma}(\sigma_s)=0, \hspace{1cm}
\frac{\partial^2 V_{eff}}{\partial \sigma^2}(\sigma_s)>0.
    \label{eq:condition}
\end{equation}
These conditions are  satisfied for wide range of parameters in our
model. 
To show it, we first analyze the potential under some approximation.
In naive analysis, we ignore the fermion contribution and 
 assume that ${\cal C}_B$ is constant because
the dependence of variables $a$ and $\sigma$ is logarithmic.
In this case, the condition (\ref{eq:condition}) is given as
%
\begin{eqnarray}
&&  m_\Lambda^2-16m_{b_0}^2 e^{-2\kappa_\sigma \sigma_s}+4{\cal C}
m_{b_0}^5 e^{-10\kappa_\sigma \sigma_s}=0
    \label{eq:condition1}
\nonumber \\
&&5m_\Lambda^2-128m_{b_0}^2 e^{-2\kappa_\sigma \sigma_s}+80{\cal C}
m_{b_0}^5 e^{-10\kappa_\sigma \sigma_s}>0,\nonumber\\
    \label{eq:condition2}
\end{eqnarray}
where we introduce the dimensionless mass scales
$m_\Lambda=M_\Lambda/M_\kappa=\sqrt{\Lambda}\kappa^{1/3}$ and $m_{b_0}
=M_{b_0}/M_\kappa=\kappa^{1/3}/b_0$, which denote
the ratios of the mass scale of a cosmological constant and that of 
internal space to the  5-dimensional Planck mass, respectively.
Introducing Eq.(\ref{eq:condition1}) into Eq.(\ref{eq:condition2}),
we find the condition
%
\begin{eqnarray}
{\cal C}_Bm_{b_0}^3 e^{-8\kappa_\sigma \sigma_s}>4.
\label{eq:condition3}
\end{eqnarray}
The value of the potential at minimal point $\sigma=\sigma_s$ is
then found to be
%
\begin{eqnarray}
V_{eff}(\sigma_s)&=&M_\kappa^5e^{-{10\over 3}\kappa_\sigma \sigma_s}
\left[
m_\Lambda^2-10m_{b_0}^2 e^{-2\kappa_\sigma \sigma_s}
\nonumber \right.
\\
&&\left.~~~~~~~~~~~~~~~~~~~~~~~~+{\cal C}_B
m_{b_0}^5 e^{-10\kappa_\sigma \sigma_s}
\right]
\nonumber \\
&=&3M_\kappa^5
m_{b_0}^2e^{-{16\over 3}\kappa_\sigma \sigma_s}
\left[2-
{\cal C}_Bm_{b_0}^3 e^{-8\kappa_\sigma \sigma_s}
\right]\nonumber \\
&<& -6M_\kappa^5
m_{b_0}^2e^{-{16\over 3}\kappa_\sigma \sigma_s}.
\label{condition_AdS}
\end{eqnarray}
In the last inequality, we have used (\ref{eq:condition3}).
This result means that the stable minimum of the potential, if it exists,
is always negative.

\begin{figure}
\includegraphics{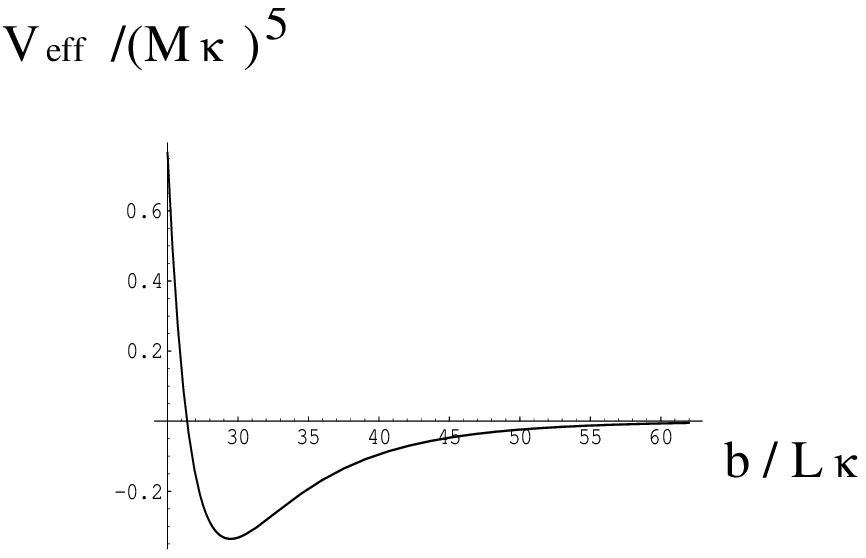}
\caption{The dilaton effective potential for $dS_5\times S^5$
 background given in Eq.(\ref{effective_pot_dS}) is 
depicted. We set $\bar{\Lambda}=1.59992\times10^{-3}M_{\kappa}^5, N_\phi=1,N_V=5, N_f=2, b_0=10^2L_{\kappa}, \mu=b_0^{-1}$ and $a=4.91490\times 10^{6}L_{\kappa}$.
The  minimum of the  potential 
is located at $\sigma_s = -5.11012 L_
\kappa$ ($b_s= b_0\exp(\kappa_{\sigma}\sigma_s)=29.4772 L_\kappa$) and 
its value
is $V_{eff}(\sigma_s) =-0.33577 M_{\kappa}^5$. 
$L_\kappa$ and $M_{\kappa}$ denote the five dimensional Planck 
length and mass, respectively.}
\label{fig:1}
\end{figure}

The above naive analysis is confirmed numerically.
Including the dependence of $\sigma$ in ${\cal C}_B$, we 
survey the parameter space and find that
the potential minimum is always negative when we have a 
stable minimum point.
For example, choosing
$N_\phi=1, N_V=5, N_f=2$, and setting
$\bar{\Lambda}=1.59992\times 10^{-3}M_{\kappa}^5, b_0=10^2 L_{\kappa},
 \mu=b_0^{-1}$, and $a=4.91490\times 10^6 L_{\kappa}$,
we find the minimum value $V_{eff}(\sigma_s)=-0.33577 M_{\kappa}^5$ 
at $\sigma_s = -5.11012 L_\kappa$
for $dS_5\times S^5$ background,
while we obtain the minimum value 
$V_{eff}(\sigma_s)=-0.30375 M_{\kappa}^5$ at
 $\sigma_s = -5.03658 L_\kappa$
for $AdS_5\times S^5$ background.
The potential form is given in Fig.\ref{fig:1}.

The universe may be created one the top of the potential hill.
Then inflation will take place.
However, such state  is not stable in the  $dS_5\times S^5$ background.
The universe rolls down to the potential minimum, which value is negative.
Then we have to switch the effective potential calculated in
 $AdS_5\times S^5$
background. In this case, we also have a negative potential minimum.
Then the universe evolves into the static $AdS_5\times S^5$, which is 
a stable spacetime.  The parameter $b_0$ is the scale of
renormalization in the quantum correction. The ratio of the parameters 
$(b_0/a)^4$ must be
 much smaller than unity in order to be renormalizable. 
 
From the consistency condition, we 
find that $V_{eff}(\sigma_s)=-0.30375 M_{\kappa}^5$ 
at $\sigma_s = -5.03658 L_{\kappa}$ for 
$a=4.91490\times 10^{6}L_{\kappa}$.
We have chosen the same values as the above ones for  $N_\phi=1,
 N_V=5, N_f=2, \bar{\Lambda}=1.59992\times 10^{-3}M_{\kappa}^5,
 b_0=10^2L_{\kappa}$, and $\mu=b_0^{-1}$.
The obtained stable $AdS_5\times S^5$ has a large scale length for 
$AdS_5$
because of inflation, while the radius of $S^5$ stays to be small.

\section{Dilaton dynamics in Type IIB supergravity model}
       \label{sec:IIBSG}
Now we consider ten dimensional type IIB supergravity
model. This model has been discussed by many authors from the motivation
of AdS/CFT correspondence\cite{ma}.     
Stelle {\it et.al}\cite{du} 
studied type IIB supergravity in $AdS_5\times
S^5$ as the candidate for the RS brane world model.   
However, the stability of $S^5$ and its dynamics in their
suggestion was not shown. 
In order to construct a realistic RS
 model, we should treat the scale of $S^5$ as
a dynamical variable
and show that it stays at small value compared with that of
$AdS_5$. 

In the cosmological point of view, de Sitter spacetime
is more interesting than anti-de Sitter spacetime because the de Sitter
spacetime corresponds to  a geometry of inflationary era in the cosmology.
In fact, the brane world cosmology in $dS_5$ spacetime
 is discussed by
several authors with the issues of inflation. In the following, we
investigate the stabilization of
$S^5$ in the $(A)dS_5\times S^5$ spacetime.          

Let us consider ten dimensional type IIB supergravity action
\cite{superstring}:
%
\begin{eqnarray}
S_{IIB}&=&S_{NS}+S_{R}+S_{CS},
      \nonumber\\
S_{NS}&=&\frac{1}{2\bar{\kappa}^2}\int d^{10}X\;\sqrt{-\tilde{g}}\;
         e^{-2\Phi}
\nonumber \\
&& \times
\left(\tilde{R}+4\tilde{g}^{MN}\partial_M\Phi\partial_N\Phi
         -\frac{1}{2}|H_3|^2\right),
      \nonumber\\
S_{R}&=&-\frac{1}{4\bar{\kappa}^2}\int d^{10}X\;\sqrt{-\tilde{g}}\;
        \left(|F_1|^2+|\tilde{F}_3|^2+\frac{1}{2}|\tilde{F}_5|^2\right),
      \nonumber\\
S_{CS}&=&-\frac{1}{4\bar{\kappa}^2}\int d^{10}X\;
          \sqrt{-\tilde{g}}\;C_4\wedge H_3\wedge F_3, 
          \label{eq;3-1}
\end{eqnarray}
where
%
\begin{eqnarray}
\tilde{F}_3&=&F_3-C_3\wedge H_3,
      \nonumber\\
\tilde{F}_5&=&F_5-\frac{1}{2}C_2\wedge H_3
         +\frac{1}{2}B_2\wedge F_3,
          \label{eq;3-2}
\end{eqnarray}
and $\tilde{R}$ is a Ricci scalar respect to the metric $\tilde{g}_{MN}$.
Note that the subscript numbers in $H_3, F_1, F_3, F_5$ etc denote
the rank of tensor fields.  
If we assume the $F_1=F_3=H_3=0$, the action is given by
%
\begin{eqnarray}
S_{IIB}&=&\frac{1}{2\bar{\kappa}^2}\int d^{10}X\sqrt{-\tilde{g}}
  \nonumber \\
&\times&\left\{e^{-2\Phi}\left(\tilde{R}+4\tilde{g}^{MN}
  \partial_M\Phi\partial_N\Phi
  \right)-\frac{1}{4}|F_5|^2\right\}.
          \label{eq;3-3}
\end{eqnarray}
Since we   consider the case when the Neveu-Schwarz(NS) scalar
 field $\Phi$ is time
dependent, the factor $e^{2\Phi}$ before the Ricci scalar $\tilde{R}$
and the kinetic term $\tilde{g}^{MN}\partial_M\Phi\partial_N\Phi$ of the
$\Phi$ is not constant. Hence the action of NS fields looks quite
different from the usual massless scalar field. Then we perform
the following conformal
 transformation 
%
\begin{eqnarray}
\tilde{g}_{MN}=e^{\Phi/2}\bar{g}_{MN}.
          \label{eq;3-4}
\end{eqnarray}
Note that we calculate the quantum effect at the local minimum of
potential $V(\Phi)$(i.e. $\Phi=0$) which arises after the 
compactification of $S^5$. Then the conformal factor becomes
unity after $\Phi$ settles to $\Phi=0$.  Consequently the change of
the frame does not affect any result, provided that our
results are interpreted after $\Phi$ settles to $\Phi=0$. 
 Using the conformal transformation (\ref{eq;3-4})
and rescaling $\Phi$ as $\Phi\rightarrow \Phi/(2\sqrt{17})$, 
the action is given by 
%
\begin{equation}
S=\frac{1}{2\bar{\kappa}^2}\int d^{10}X\sqrt{-\bar{g}}
  \left\{\left(\bar{R}
  -\bar{g}^{MN}\partial_M\Phi\partial_N\Phi
  \right)-\frac{1}{4}|F_5|^2\right\},
          \label{eq;3-5}
\end{equation}
where $\bar{R}$ is a Ricci scalar of the ten dimensional metric tensor
$\bar{g}_{\mu\nu}$.  
The background metric is expressed as 
Eq.(\ref{eq;2-2})\footnote{Although the supersymmetry
 is broken in this compactification\cite{pi}, 
there is no problem in the cosmological point of
 view.}.
We consider the Freund-Rubin type solution as 5-form
field strength; 
%
\begin{eqnarray}
F_{M_1\cdots M_5}&=&\left\{ 
\begin{array}{ll}
 \left(f/\sqrt{\Omega^{(5)}}\right)\epsilon_{M_1\cdots M_5},& 
M_1=i, \cdots, M_5=m,\\
 0, & {\rm otherwise},
\end{array}
\right.\nonumber\\
          \label{eq;3-6}
\end{eqnarray}
where $f$ is a constant and $\sqrt{\Omega^{(5)}}$ is
 the determinant of five dimensional 
sphere $S^5$.
This field strength is wrapped around the $S^5$.
We then set $|F_5|^2\equiv 8\bar{\Lambda}$ and
redefine the NS scalar $\varphi=\Phi/\bar{\kappa}$.
 The ten dimensional action (\ref{eq;3-5}) is now 
rewritten by 
%
\begin{eqnarray}
S=\int d^{10}X\sqrt{-\bar{g}}
  \left[\frac{1}{2\bar{\kappa}^2}\left(\bar{R}-2\bar{\Lambda}\right)
 -\frac{1}{2}
    \bar{g}^{MN}\partial_M\varphi\partial_N\varphi 
         \right].\nonumber\\
\label{eq;3-5-1}
\end{eqnarray}
To write down the five dimensional effective action,
the NS scalar $\varphi$ is expanded as 
%
\begin{equation}
 \varphi = b_{0}^{-5/2}
	\sum_{l,m}\varphi_{lm}(x^{\mu})Y^{(5)}_{lm}(y^i),
        \label{eq;pharmo}
\end{equation}
where $b_0$ is the initial value of $b$. 
Substituting the metric (\ref{eq;2-2})and the NS scalar (\ref{eq;pharmo}) into 
the action (\ref{eq;3-5-1}), we get the five dimensional effective action 
%
\begin{eqnarray}
&&S=\int d^{5}x\sqrt{-g}
  \left[
\left(\frac{1}{2\kappa^2}R
  -\frac{1}{2}g^{\mu\nu}\partial_{\mu}\sigma\partial_{\nu}\sigma
  -U(\sigma)\right)
\right. \nonumber \\
&&
\left.
~ -\sum_{l,\;m}\frac{1}{2}
   \left(g^{\mu\nu}\partial_{\mu}
   \varphi_{lm}\partial_{\nu}\varphi_{lm}
   +M^2_{\varphi}\varphi_{lm}^2\right)
\right]  .
        \label{eq;3-10}
\end{eqnarray}
where $R$ is Ricci scalar of five dimensional metric tensor
$g_{\mu\nu}$, $\kappa$ is a positive constant defined by
$\kappa^2=\bar{\kappa}^2/(2^{5}b_0^{5}\pi)$,
$\sigma$ is given by (\ref{eq;2-4}), the mass $M^2_{\varphi}$
 of five dimensional NS scalar field
$\varphi$ is given by
%
\begin{equation} 
M^2_{\varphi}=\frac{l(l+4)}{b_0^2}e^{-{16\over 3}\kappa_\sigma\sigma}, 
\end{equation}
and the dilaton potential $U(\sigma)$ is given by
%
\begin{equation}
U(\sigma)=
\frac{\bar{\Lambda}}{\kappa^2}e^{-{10\over 3}\kappa_\sigma\sigma}
        -\frac{10}{\kappa^2 b_0^2}
e^{-{16\over 3}\kappa_\sigma\sigma}.
          \label{eq;d-potential}
 \end{equation}
Note that the classical NS scalar fields $\varphi_{lm}$ in the action (\ref{eq;3-10}) 
has a potential, which vanishes at $\varphi_{lm}=0$, unless $l=m=0$  
which does not couple to the dilaton. 
Furthermore, the kinetic term of $\varphi$
is also approximately zero because the NS scalar is stable at
 $\varphi_{lm}=0~(l\neq 0)$. 
Then we expect that  the massive scalar field $\varphi_{lm}=0~(l\neq 0)$ 
has a zero vacuum expectation value
 at least classical level, resulting in that 
this scalar field does not produce the energy momentum except for the massless mode.   
This background geometry has 3-brane which act as source for the 4-form
field.

Here we consider the quantum effect of NS scalar. The NS scalar denotes the
scale of eleventh dimension in eleven dimensional supergravity, which is compactified 
a la
Kaluza-Klein. This length scale is not too far above
Planck length. Even though a satisfactory quantum theory of gravity is not
known so far, we expect that the quantum effect is presumably very important.       
As the action (\ref{eq;3-10}) is same form as (\ref{eq;2-3}) and 
(\ref{eq;2-8}),  
the 1-loop effective potential for NS scalar is given by  
%
\begin{eqnarray}
&&V_{eff}(\sigma)=\frac{\bar{\Lambda}}{\kappa^2}
e^{-{10\over 3}\kappa_\sigma\sigma}
        -\frac{10}{\kappa^2 b_0^2}
e^{-{16\over 3}\kappa_\sigma\sigma}
       +\frac{1}{b_0^5}
        e^{-{40\over 3}\kappa_\sigma\sigma}
 \nonumber \\
&&~
\times\left[-\ln\left(\mu^2\:a^2\right)+
        C\ln \left\{\left(\frac{a}{b_0}\right)^2    
e^{-{16\over 3}\kappa_\sigma\sigma}
\right\}\right],
          \label{eq;3-11}
\end{eqnarray}
where $C=25/(8192\pi)$ 
(See Eq.(\ref{eq;3-constant})).   
This is exactly same as that calculated in \S\ref{sec:stabilization}.
We have also other fields which give the contribution
to the effective potential.
Such a contribution has also been calculated in \S\ref{sec:stabilization}.
In the background is $AdS_5\times S^5$,
we again find Eqs.(\ref{QC_AdS}) and (\ref{QC2_AdS}).
 
 In the present model, we have $N_\phi=1, N_V=5,$ and $N_f=2$.
Setting $\bar{\Lambda}=1.59992\times 10^{-3}M_{\kappa}^5,
 b_0=10^2L_{\kappa}$ and
 $\mu=b_0^{-1}$, we find
the stable  minimum point $\sigma_s=-5.03658 L_{\kappa}$ where 
$V_{eff}(\sigma_s)=-0.30375 M_{\kappa}^5$
 for $a=4.91490\times 10^{6}L_{\kappa}$.
We may conclude that the stable RS model is realized from Type IIB
supergravity theory.

We may also have to include further contributions from other modes of
gravitons and gravitinos, which are ignored in the present analysis
because of a technical reason. It might be 
justified from the following reason.
If supersymmetry is not broken,
quantum correction should vanish at the $AdS_5$ minimum and then a stable
compactification is not obtained.
 Hence, here,
we assume that supersymmetry is broken by an unknown mechanism in order to
find a stable minimum.

\section{Conclusion}
\label{sec:conclusion}
We have calculated quantum effects  in the $(A)dS_5\times S^5$
compactified background of  the ten dimensional 
Kaluza-Klein theory and type IIB supergravity 
and discussed its stability using the effective potential.  
In their pure gravity systems,  
a curvature term of the internal space gives a dominant contribution
to the dilaton potential at small $b$, while a cosmological constant term
becomes dominant for large $b$. 
Hence the dilaton potential is  unbounded from below as $b\rightarrow 0$ 
and drops exponentially as $b\rightarrow \infty$.
 Then the extra
 dimension either shrinks to zero volume
or is   decompactified.
However, if we include quantum effects,
we find a stable minimum for the dilaton potential. 
 In the ten dimensional Kaluza-Klein model, the cosmological constant and
 quantum correction of various matter fields
 force to expand the extra dimension while 
the curvature of the internal spacetime forces to contract it. 
These combination produces a local minimum of the effective potential.
In ten dimensional type IIB supergravity model,
the scale of extra dimension is stabilized by balancing the five form 
gauge field strength wrapped around the $S^5$,
the curvature term of $S^5$ and quantum correction term induced by the NS 
scalar.    The NS scalar is originally characterized by the direction 
of eleventh dimension in the eleven dimensional supergravity.   
Thus the quantum effect of NS scalar becomes important if the
eleventh dimension is compactified near the Planck scale. 

When the universe is created, the dilaton may be located  near  the top
of the potential hill, 
 and then the background geometry is almost $dS_5$ spacetime. 
We then have exponential expansion of the 5-dimensional spacetime.
As the dilaton rolls down the potential hill, 
the five dimensional spacetime geometry deviates from
the $dS_5$. The dilaton  potential eventually turns out to be 
 negative, and the dilaton finds a stable minimum point. Then, the  five
dimensional spacetime becomes 
$AdS_5$.  Hence, a five dimensional de Sitter ($dS_5$) spacetime evolves
 into a five dimensional anti-de Sitter($AdS_5$) when 
 the dilaton settles down to the potential minimum.
Associated with dilaton
 stabilization, the change of spacetime geometry takes place.

Our solution contains a 3-brane because the background geometry has the
five form gauge field strength. Hence, 
the ten dimensional $(A)dS_5\times S^5$ compactification may provide  a
realistic model of the RS brane world\cite{ra}.

However, if we have branes in our $AdS_5$, we may find 
additional contributions of quantum effect because
we have new boundaries of branes.
It may change some part of the present results, although we believe
that $AdS_5$ is still stable.
We leave this calculation as a future work.

\begin{acknowledgments}
We would like to thank T. Kubota, O. Yasuda, N. Sakai, and T. Torii 
for valuable comments.
KU is also grateful
to M. Sakagami and J. Soda for 
continuing encouragement.
This work was partially supported by the Grant-in-Aid  for Scientific
Research  Fund of the Ministry of Education, Science and Culture (Nos.
14047216, 14540281) and by the Waseda University Grant for Special
Research Projects.
\end{acknowledgments}
\appendix
\section{Zeta function regularization in $AdS_n\times S^d$}
\subsection{scalar field}
\label{sec:scalar}
\begin{figure}
\includegraphics{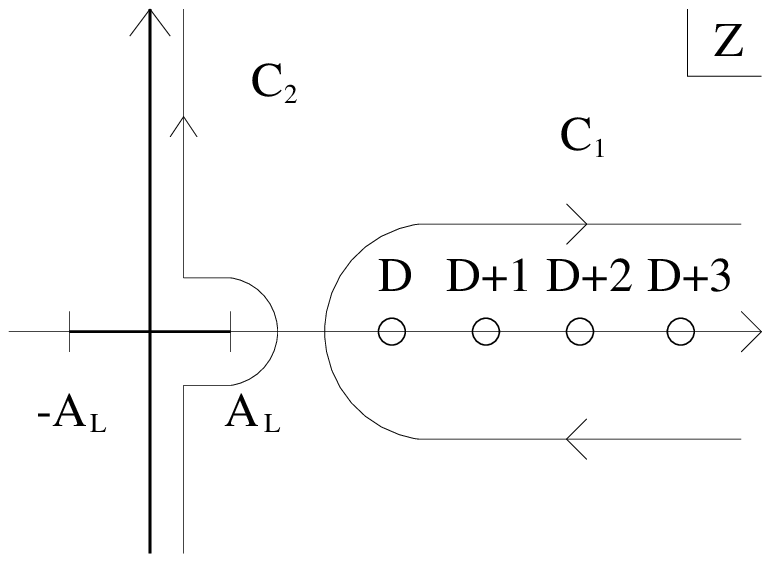}
\caption{The contour $C_1$ in Eq.(\ref{eq;a-6}) is replaced by the
 contour $C_2$. Note that the contour $C_2$ avoid the branch point at
 $z=\pm A_L$.}
\label{fig:s1}
\end{figure}

In this Appendix, we provide the method to calculate the zeta
function regularization for scalar field in the product spacetime 
 $AdS_n\times S^d$. 
The Euclidean section for $AdS_n$ spacetime is the
$n$-dimensional hyperbolic space $H^n$. The calculation of zeta function
for $AdS_n$ is discussed in \cite{cam1}. We extend their
calculation technique to the zeta function for $AdS_n\times S^d$.    
On the compact Euclidean section, the zeta function is given by\cite{uz1} 
%
\begin{equation}
\zeta_{\phi}(s)=\sum^{\infty}_{l=0}D_l\;\Lambda_l^{-s}, 
\end{equation}
where $\Lambda_l$ is the discrete eigenvalue of the Laplace-Beltrami
operator and $D_l$ is the degeneracy of the eigenvalue.
The calculation of zeta function on the $S^d$ is  performed using
well-known spectrum of the Laplace-Beltrami operator on $S^d$.
On the other hand, the zeta function for the noncompact manifold is not
same as that in a compact case. On the homogeneous $n$-dimensional
hyperbolic space $H^n$, the zeta function takes the form\cite{cam1}
%
\begin{equation}
\zeta_{\phi}(s)=\int^{\infty}_0 d\lambda\;\mu(\lambda)\; 
               \Lambda(\lambda)^{-s}, 
\end{equation}
where $\Lambda(\lambda)$ is the eigenvalue of the Laplace-Beltrami
operator on the $H^n$ and $\lambda$ is the real parameter which labels
the continuous spectrum and $\mu(\lambda)$ is the spectrum function (or
Plancherel measure) on $H^n$ corresponding to the discrete degeneracy on
the $S^d$. 
The spectrum function for the scalar and spinor fields
on the $H^n$ was already  calculated in \cite{cam2,cam3}. The
spectrum function for the transverse-traceless tensor field 
on $H^n$ was also given in \cite{cam4}.
Define the generalized zeta function in $H^n\times S^d$
for the scalar field as  
%
\begin{eqnarray}
\zeta_{\phi}(s)&=&\sum^{\infty}_{l=0}
         \frac{(l+d-2)!}{(d-1)!}
         \frac{(2l+d-1)}{l!}
         \int^{\infty}_0 d\lambda\;\mu(\lambda)
            \nonumber\\
        & & \hspace{0.5cm}
            \times\left\{\frac{\lambda^2+\left\{(n-1)/2\right\}^2}{a^2}
            \right.\nonumber\\
        & & \left.+\frac{l(l+d-1)}{b_0^2}
                     e^{-[2d/(n-2)+2]\kappa_\sigma \sigma}\right\}^{-s}.
       \label{eq;a-1}
\end{eqnarray}
For the odd dimension,
Plancherel measure $\mu(\lambda)$ is given by\cite{cam1}
%
\begin{equation}
\mu(\lambda)=\frac{\pi}{2^{2(n-2)}\left\{\Gamma(n/2)\right\}^2}
             \prod^{(n-3)/2}_{j=0}\left(\lambda^2+j^2\right).
       \label{eq;a-2}
\end{equation}
Using $D=(d-1)/2$, $N=(n-1)/2$ instead of $d$, $n$ and running variables
$L=l+D$, we rewrite (\ref{eq;a-1}) as  
%
\begin{equation}
\zeta_{\phi}(s)=\sum^{\infty}_{L=D}D_{\phi}(L)\int^{\infty}_0 
           d\lambda\;\mu(\lambda)
         \left(\frac{\lambda^2+N^2}{a^2}
         +M^2\right)^{-s}, 
         \label{eq;a-3}
\end{equation}
where the $D_{\phi}(L)$, $\Lambda_{\phi}(L)$ are given by
%
\begin{eqnarray}
D_{\phi}(L)&=&\frac{2L^2}{(2D)!}\left\{L^2-(D-1)^2\right\}
            \cdots\left\{L^2-1\right\},\nonumber\\
\Lambda_{\phi}(L)&=&L^2-D^2,\nonumber\\
M_{\phi}^2&=&\frac{\Lambda_{\phi}(L)}{b_0^2}
                 e^{-[2d/(n-2)+2]\kappa_\sigma \sigma}.
        \label{eq;a-4}
\end{eqnarray}
Integrating Eq.(\ref{eq;a-3}) with respect to
 $\lambda$, we find the expression   
%
\begin{eqnarray}
\zeta_{\phi}(s)&=&\frac{\pi^{3/2}a^{2s}}{2^{2n-1}
       \left\{\Gamma(2N+1)\right\}^2}
        \frac{1}{\Gamma(s)}\sum^{\infty}_{L=D}D_{\phi}(L)
        \nonumber\\ 
        &\times&\left[3\left(N^2+a^2 M^2\right)^{-s+5/2}
       \Gamma\left(s-\frac{5}{2}\right)\right.
         \nonumber\\
       & &\left.+2\left(N^2+a^2 M^2\right)^{-s+3/2}
        \Gamma\left(s-\frac{3}{2}\right)\right].
        \label{eq;a-5}
\end{eqnarray}
To regularize the mode sum in Eq.(\ref{eq;a-5}), we replace the infinite
sum for $L$ by complex integration.
The generalized zeta function is 
%
\begin{widetext}
\begin{eqnarray}
\zeta_{\phi}(s)&=&\frac{\pi^{3/2}a^{2s}}
       {2^{2n-1}\left\{\Gamma(2N+1)\right\}^2\Gamma(s)}
        \nonumber\\ 
        &\times&\left[\frac{3i}{2}
       \left\{\left(\frac{a}{b_0}\right)^2
        e^{-[2d/(n-2)+2]\kappa_\sigma \sigma}
\right\}^{-s+5/2}
       \int_{C_1}dz\;D_{\phi}(z)\:\cot(\pi z)
       \left(z^2-A_L^2\right)^{-s+5/2}
       \Gamma\left(s-\frac{5}{2}\right)\right.\nonumber\\
       & + &\left. i\left\{\left(\frac{a}{b_0}\right)^2
        e^{-[2d/(n-2)+2]\kappa_\sigma \sigma}
\right\}^{-s+3/2}
       \int_{C_1}dz\;D_{\phi}(z)\:\cot(\pi z)
       \left(z^2-A_L^2\right)^{-s+3/2}
        \Gamma\left(s-\frac{3}{2}\right)\right],
        \label{eq;a-6}
\end{eqnarray}
\end{widetext}
where  $A^2_L$ is given by

%
\begin{equation}
A_L^2=D^2
      -N^2\left(\frac{b_0}{a}\right)^2
              e^{[2d/(n-2)+2]\kappa_\sigma \sigma}.
       \label{eq;a-7}
\end{equation}
and the contour $C_1$ in the complex plane is showed 
in Fig.\ref{fig:s1}.
Note that there are two branch points $z=\pm A_L$ in the integration.
We introduce the function $\tilde{\zeta}_k(s)$ given by
%
\begin{eqnarray}
& &\tilde{\zeta}_k(s) =
        \frac{\pi^{3/2}a^{2s}}{2^{2n-1}
        \left\{\Gamma(2N+1)\right\}^2\Gamma(s)}
        \nonumber\\
        &\times&
        \left[\frac{i}{2}
       \left\{\left(\frac{a}{b_0}\right)^2
        e^{-[2d/(n-2)+2]\kappa_\sigma \sigma}
       \right\}^{-s+k/2}
       \right.\nonumber\\ 
        &\times&\left.\int_{C_1}dz\;D_{\phi}(z)\:\cot(\pi z)
       \left(z^2-A_L^2\right)^{-s+k/2}
       \Gamma\left(s-\frac{k}{2}\right)\right].
        \nonumber\\
        \label{eq;a-8}
\end{eqnarray}
Using the above definition, $\zeta_{\phi}(s)$ is then expressed by 
%
\begin{equation}
\zeta_{\phi}(s)=3\:\tilde{\zeta}_5(s)+2\:\tilde{\zeta}_3(s).
       \label{eq;a-9}
\end{equation}
 Now we move the
contour $C_1$ to the parallel line along the imaginary axis in order
to account the poles for $\cot(\pi z)$ in (\ref{eq;a-8}) (See Fig.\ref{fig:1}).
The contour 
$C_2$ is replaced with  lines passing just above the cuts associated with $z=\pm A_L$.
$\tilde{\zeta}(s)$ is then given by 
%
\begin{widetext}
\begin{eqnarray}
\tilde{\zeta}_k(s)&=&
       \frac{\pi^{3/2}a^{2s}}{2^{2n-1}
    \left\{\Gamma(2N+1)\right\}^2\Gamma(s)}
       \left\{\left(\frac{a}{b_0}\right)^2
              e^{-[2d/(n-2)+2]\kappa_\sigma \sigma}
\right\}^{-s+k/2}
      \sin\left\{\pi\left(s-\frac{k}{2}\right)\right\}\nonumber\\
      & \times &\left\{\int^{\infty}_0dx\;\:D_{\phi}(ix)
      \left(x^2+A_L^2\right)^{-s+k/2}\coth(\pi x)
     -\int^{A_L}_0
      dx\:\cot(\pi x)\;D_{\phi}(x)\left(A_L^2-x^2\right)^{-s+k/2}
      \right\},
       \label{eq;a-10}
\end{eqnarray}
\end{widetext}
where $D_{\phi}(ix)$ is the polynomial with coefficients $r_{Nk}$;
%
\begin{eqnarray}
D_{\phi}(ix)&=&(-1)^{D}\frac{2x^2}{(2D)!}
        \left\{x^2+\left(D-1\right)^2\right\}\cdots
        \left\{x^2+1\right\}\nonumber\\
       &\equiv&(-1)^{D}\sum^{D-1}_{p=0}\;r_{Np}\;x^{2p+2}.
       \label{eq;a-11}
\end{eqnarray}
The first term in Eq.(\ref{eq;a-8}) comes from the integral 
along the imaginary
axis and the  second  term in Eq.(\ref{eq;a-8}) 
is contribution from the
contours along the cuts of on the real axis, respectively.
Substituting  the relation 
%
\begin{equation}
\coth(\pi x)=1+\frac{2}{e^{2\pi x}-1},
       \label{eq;a-12}
\end{equation}
 into the first term of (\ref{eq;a-10}), 
the function $\tilde{\zeta}_k(s)$ is finally given by  
%
\begin{widetext}
\begin{eqnarray}
\tilde{\zeta}_k(s)&
            =&\frac{\pi^{3/2}a^{2s}}{2^{2n-1}
         \left\{\Gamma(2N+1)\right\}^2
           \Gamma(s)}
          \left\{\left(\frac{a}{b_0}\right)^2
         e^{-[2d/(n-2)+2]\kappa_\sigma \sigma}
\right\}^{-s+k/2}
        \sin\left\{\pi\left(s-\frac{k}{2}\right)\right\}\nonumber\\
      & &\times      
      \left\{\frac{1}{2}\sum^{D-1}_{p=0}r_{Np}
      \left(A_L^2\right)^{p+4-s}\frac{\Gamma\left(p+\frac{3}{2}\right)
      \Gamma(s-p-4)}
      {\Gamma\left(s-\frac{k}{2}\right)}\right.
      \nonumber\\
    & &-\left.\int^{\infty}_0 dx\;D_{\phi}(ix)
       \left(x^2+A_L^2\right)^{-s+k/2}
       \frac{2}{e^{2\pi x}-1}+\int^{A_L}_0 dx D_{\phi}(x)
        \left(A_L^2-x^2\right)^{-s+k/2}\cot(\pi x)\right\}.
         \label{eq;a-13}
\end{eqnarray}
\end{widetext}

\subsection{spinor field}
         \label{sec:spinor}
\begin{figure}
\includegraphics{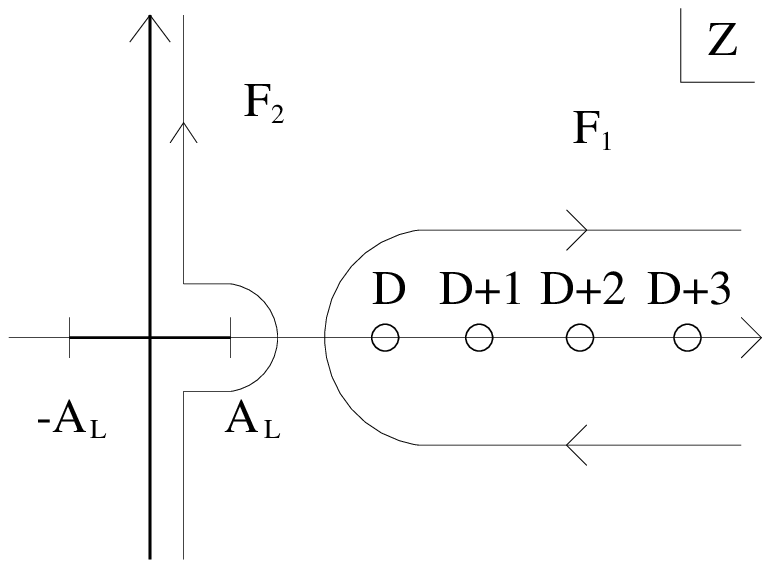}
\caption{The contour $F_1$ in Eq.(\ref{eq;b-6}) is replaced by the
 contour $F_2$. Note that the contour $F_2$ avoid the branch point at
 $z=\pm A_L$.}
\label{fig:f1}
\end{figure}

Here we discuss the zeta function regularization 
 for the spin 1/2 field
in $H^n\times S^d$ when both $n$ and $d$ are odd. 
We define the generalized zeta function in $H^n\times S^d$
for the spinor field as follows 
%
\begin{eqnarray}
\zeta_f(s)&=&\sum^{\infty}_{l=0}
         \frac{\Gamma\left(l+\frac{d}{2}\right)}{\Gamma(d)\;l!}
         \int^{\infty}_0 d\lambda\;\mu(\lambda)
            \nonumber\\
          & &\times \left\{\frac{\lambda^2
            -\frac{n(n-1)}{4}}{a^2}
            +\frac{\left(l+\frac{d}{2}\right)}{b^2}\right\}^{-s},
       \label{eq;b-1}
\end{eqnarray}
where the spectral function $\mu(\lambda)$ in odd dimension 
is given by\cite{cam3}
%
\begin{equation}
\mu(\lambda)=\frac{\pi}{2^{2(n-2)}\left\{\Gamma(n/2)\right\}^2}
             \prod^{(n-2)/2}_{j=1/2}\left(\lambda^2+j^2\right).
       \label{eq;b-2}
\end{equation}
Using $D=d/2$, $N=n/2$ instead of $d$, $n$ and running variables
$L=l+D$, we rewrite (\ref{eq;b-1}) as  
%
\begin{eqnarray}
\zeta_f(s)&=&
          \sum^{\infty}_{L=D}D_f(L)\int^{\infty}_0 d\lambda\;\mu(\lambda)
          \nonumber\\
          & &\times 
         \left\{\frac{\lambda^2-N\left(N-\frac{1}{2}\right)}{a^2}
         +M^2\right\}^{-s}, 
         \label{eq;b-3}
\end{eqnarray}
where the $D_f(L)$, $\Lambda_f(L)$ are given by
%
\begin{eqnarray}
D_f(L)&=&\frac{1}{(2D)!}\left\{L^2-(D-1)^2\right\}
            \cdots\left\{L^2-\frac{1}{4}\right\},\nonumber\\
\Lambda_f(L)&=&L^2-D^2,\nonumber\\
M_f^2&=&\frac{\Lambda_f(L)}{b^2}.
        \label{eq;b-4}
\end{eqnarray}
Integrating Eq.(\ref{eq;b-3}) with respect to
 $\lambda$, we find the expression   
%
\begin{widetext}
\begin{eqnarray}
\zeta_f(s)&=&\frac{\pi^{3/2}a^{2s}}{2^{2n-1}\left\{\Gamma(2N)\right\}^2}
        \frac{1}{\Gamma(s)}\sum^{\infty}_{L=D}D_f(L)
       \left[12\left\{N\left(N-\frac{1}{2}\right)+a^2 M^2\right\}^{-s+5/2}
       \Gamma\left(s-\frac{5}{2}\right)\right.\nonumber\\
       & &\left.
       +20\left\{N\left(N-\frac{1}{2}\right)+a^2 M^2\right\}^{-s+3/2}
        \Gamma\left(s-\frac{3}{2}\right)
       +9\left\{N\left(N-\frac{1}{2}\right)
        +a^2 M^2\right\}^{-s+1/2}
        \Gamma\left(s-\frac{1}{2}\right)\right].\nonumber\\
        \label{eq;b-5}
\end{eqnarray}
\end{widetext}
Replacing the infinite
sum for $L$ by complex integration,
the generalized zeta function is found to be
%
\begin{widetext}
\begin{eqnarray}
\zeta_f(s)&=&\frac{\pi^{3/2}a^{2s}}
       {2^{2n-1}\left\{\Gamma(2N)\right\}^2\Gamma(s)}
        \nonumber\\ 
        &\times&\left[6i
       \left\{\left(\frac{a}{b_0}\right)^2
        e^{-[2d/(n-2)+2]\kappa_\sigma \sigma}
\right\}^{-s+5/2}
       \int_{F_1}dz\;D_f(z)\:\tan(\pi z)
       \left(z^2-A_L^2\right)^{-s+5/2}
       \Gamma\left(s-\frac{5}{2}\right)\right.\nonumber\\
       & + &10i\left\{\left(\frac{a}{b_0}\right)^2
        e^{-[2d/(n-2)+2]\kappa_\sigma \sigma}
\right\}^{-s+3/2}
       \int_{F_1}dz\;D_f(z)\:\tan(\pi z)
       \left(z^2-A_L^2\right)^{-s+3/2}
        \Gamma\left(s-\frac{3}{2}\right)\nonumber\\
       & + &\left.\frac{9i}{2}\left\{\left(\frac{a}{b_0}\right)^2
        e^{-[2d/(n-2)+2]\kappa_\sigma \sigma}
\right\}^{-s+1/2}
       \int_{F_1}dz\;D_f(z)\:\tan(\pi z)
       \left(z^2-A_L^2\right)^{-s+1/2}
        \Gamma\left(s-\frac{1}{2}\right)   \right],
        \label{eq;b-6}
\end{eqnarray}
\end{widetext}
where the contour $F_1$ in the complex plane is showed 
in Fig.\ref{fig:f1} and $A^2_L$ is given by
%
\begin{equation}
A_L^2= -N\left(N-\frac{1}{2}\right)\left(\frac{b_0}{a}\right)^2
              e^{[2d/(n-2)+2]\kappa_\sigma \sigma}.
       \label{eq;b-7}
\end{equation}
Note that there are two branch points $z=\pm A_L$ in the integration.
Define the function $\tilde{\zeta}_k(s)$;
%
\begin{eqnarray}
\tilde{\zeta}_k(s)&=&
        \frac{\pi^{3/2}\Gamma\left(s-\frac{k}{2}\right)a^{2s}}
       {2^{2n-1}\left\{\Gamma(2N)\right\}^2\Gamma(s)}
        \nonumber\\
       & \times &
        \left[\frac{i}{2}
       \left\{\left(\frac{a}{b_0}\right)^2
               e^{-[2d/(n-2)+2]\kappa_\sigma \sigma}\right\}^{-s+k/2}
       \right.\nonumber\\ 
        &\times&\left.\int_{F_1}dz\;D_f(z)\:\tan(\pi z)
       \left(z^2-A_L^2\right)^{-s+k/2}\right].
       \nonumber\\
        \label{eq;b-8}
\end{eqnarray}
The zeta function $\zeta_f(s)$ is then expressed by 
%
\begin{equation}
\zeta_f(s)=12\:\tilde{\zeta}_5(s)+20\:\tilde{\zeta}_3(s)
         +9\:\tilde{\zeta}_1(s) .
       \label{eq;b-9}
\end{equation}
 Now we move the
contour $F_1$ to the parallel line along the imaginary axis in order
to account the poles for $\tan(\pi z)$ in (\ref{eq;a-8}) (See Fig.\ref{fig:1}).
The contour 
$F_2$ is replaced with  lines passing just above the cuts associated with 
$z=\pm A_L$.
$\tilde{\zeta}_f(s)$ is then given by 
%
\begin{eqnarray}
\tilde{\zeta}_k(s)&=&
       \frac{\pi^{3/2}a^{2s}}{2^{2n-1}
         \left\{\Gamma(2N)\right\}^2\Gamma(s)}
       \sin\left\{\pi\left(s-\frac{k}{2}\right)\right\}
       \nonumber\\
       &\times&\left\{\left(\frac{a}{b_0}\right)^2
               e^{-[2d/(n-2)+2]\kappa_\sigma \sigma}
        \right\}^{-s+k/2}\nonumber\\
      & \times &\left\{\int^{\infty}_0dx\;\:D_f(ix)
      \left(x^2+A_L^2\right)^{-s+k/2}\tanh(\pi x)\right.
        \nonumber\\
      &-&\left.\int^{A_L}_0
      dx\:\tan(\pi x)\;D_f(x)\left(A_L^2-x^2\right)^{-s+k/2}
      \right\},
        \nonumber\\
       \label{eq;b-10}
\end{eqnarray}
where $D_f(ix)$ is the polynomial with coefficients $f_{Nk}$;
%
\begin{eqnarray}
D_f(ix)&=&(-1)^{(2D-1)/2}\frac{\left\{x^2+\left(D-1\right)^2\right\}
        \cdots\left\{x^2+1\right\}}{(2D)!}\nonumber\\
       &\equiv&(-1)^{(2D-1)/2}\sum^{(2D-1)/2}_{p=0}\;f_{Np}\;x^{2p}.
       \label{eq;b-11}
\end{eqnarray}
The first term in Eq.(\ref{eq;b-8} comes from the integral 
along the imaginary
axis and the  second  term in Eq.(\ref{eq;b-8}) 
is contribution from the
contours along the cuts of on the real axis, respectively.
Substituting the relation
%
\begin{equation}
\tanh(\pi x)=1-\frac{2}{e^{2\pi x}+1},
       \label{eq;b-12}
\end{equation}
 into the first term of (\ref{eq;b-10}), 
the function $\tilde{\zeta}_k(s)$ is finally given by  
%
\begin{widetext}
\begin{eqnarray}
\tilde{\zeta}_k(s)&=&
        \frac{\pi^{3/2}a^{2s}}{2^{2n-1}
        \left\{\Gamma(2N)\right\}^2\Gamma(s)}
        \sin\left\{\pi\left(s-\frac{k}{2}\right)\right\}
        \left\{\left(\frac{a}{b_0}\right)^2
        e^{-[2d/(n-2)+2]\kappa_\sigma \sigma}\right\}^{-s+k/2}
        \nonumber\\
     &\times& \left\{\frac{1}{2}\sum^{(2D-1)/2}_{p=0}f_{Np}
      \left(A_L^2\right)^{p+4-s}\frac{\Gamma\left(p+\frac{3}{2}\right)
      \Gamma(s-p-4)}
      {\Gamma\left(s-\frac{k}{2}\right)}\right.
      \nonumber\\
    &+&\left.\int^{\infty}_0 dx\;D_f(ix)
       \left(x^2+A_L^2\right)^{-s+k/2}
       \frac{2}{e^{2\pi x}+1}
       +\int^{A_L}_0 dx D_f(x)
        \left(A_L^2-x^2\right)^{-s+k/2}\tan(\pi x)\right\}.
         \nonumber\\
         \label{eq;b-13}
\end{eqnarray}
\end{widetext}
\subsection{vector field}
         \label{sec:vector}
\begin{figure}
\includegraphics{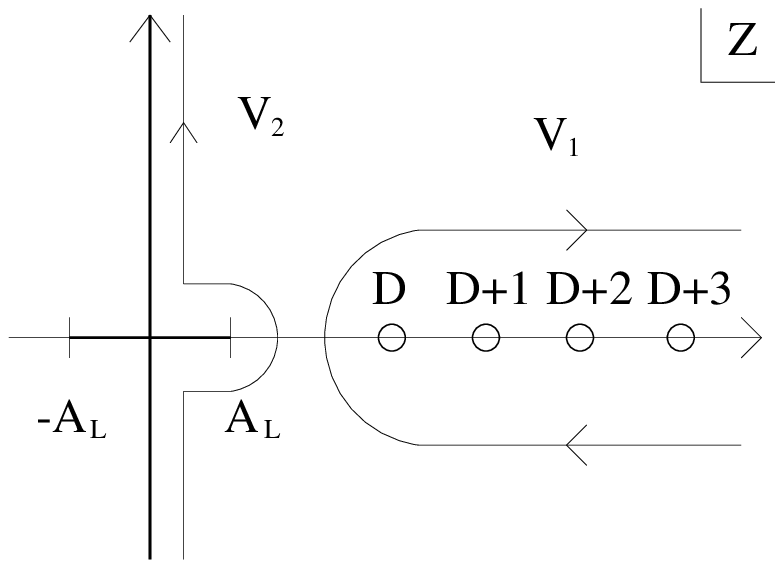}
\caption{The contour $V_1$ in Eq.(\ref{eq;a-6}) is replaced by the
 contour $V_2$. Note that the contour $C_2$ avoid the branch point at
 $z=\pm A_L$.}
\label{fig:v1}
\end{figure}

Next, we provide  the zeta function
regularization for the transverse vector
field in the product geometry 
of $H^n\times S^d$. 
Now we define the generalized zeta function in $H^n\times S^d$
for the transverse vector field as 
%
\begin{eqnarray}
\zeta_V(s)&=&\sum^{\infty}_{l=0}
         \frac{(l+d-2)!}{(d-1)!}
         \frac{(2l+d-1)}{l!}
         \int^{\infty}_0 d\lambda\;\mu(\lambda)
            \nonumber\\
        &\times&\left\{\frac{\lambda^2+\left\{(n-1)/2\right\}^2+1}{a^2}
            \right.\nonumber\\
        &+&\left.\frac{l(l+d-1)}{b_0^2}
            e^{-[2d/(n-2)+2]\kappa_\sigma \sigma}
\right\}^{-s},
       \label{eq;c-1}
\end{eqnarray}
where $\mu(\lambda)$ is the Plancherel measure. For the odd dimension,
that is given by\cite{cam4}
%
\begin{equation}
\mu(\lambda)=\frac{\pi\left\{\lambda^2+\left(1+\frac{n-3}{2}\right)^2
             \right\}}{2^{2(n-2)}
             \left\{\Gamma(n/2)\right\}^2}
             \prod^{(n-5)/2}_{j=0}\left(\lambda^2+j^2\right).
       \label{eq;c-2}
\end{equation}
Using $D=(d-1)/2$, $N=(n-1)/2$ instead of $d$, $n$ and running variables
$L=l+D$, we rewrite (\ref{eq;c-1}) as  
%
\begin{equation}
\zeta_V(s)=\sum^{\infty}_{L=D}D_V(L)\int^{\infty}_0 d\lambda\;\mu(\lambda)
         \left(\frac{\lambda^2+1+N^2}{a^2}
         +M^2\right)^{-s}, 
         \label{eq;c-3}
\end{equation}
where the $D_V(L)$, $\Lambda_V(L)$ are given by
%
\begin{eqnarray}
D_V(L)&=&\frac{2L^2}{(2D)!}\left\{L^2-(D-1)^2\right\}
            \cdots\left\{L^2-1\right\},\nonumber\\
\Lambda_V(L)&=&L^2-D^2,\nonumber\\
M_V^2&=&\frac{\Lambda_V(L)}{b_0^2}
                 e^{-[2d/(n-2)+2]\kappa_\sigma \sigma}.
        \label{eq;c-4}
\end{eqnarray}
Integrating Eq.(\ref{eq;c-3}) with respect to
 $\lambda$, we get the expression   
%
\begin{eqnarray}
\zeta_V(s)&=&\frac{\pi^{3/2}a^{2s}}{2^{2n-1}\left\{\Gamma(2N+1)\right\}^2}
        \frac{1}{\Gamma(s)}\sum^{\infty}_{L=D}D_V(L)
        \nonumber\\ 
        &\times&\left[3\left(N^2+a^2 M^2\right)^{-s+5/2}
       \Gamma\left(s-\frac{5}{2}\right)
         \right.\nonumber\\
       &+&\left.8\left(N^2+a^2 M^2\right)^{-s+3/2}
        \Gamma\left(s-\frac{3}{2}\right)\right].
        \label{eq;c-5}
\end{eqnarray}
Replacing the infinite
sum for $L$ by complex integration, the generalized zeta function is given by
%
\begin{widetext}
\begin{eqnarray}
\zeta_V(s)&=&\frac{\pi^{3/2}a^{2s}}
       {2^{2n-1}\left\{\Gamma(2N+1)\right\}^2\Gamma(s)}
        \nonumber\\ 
        &\times&\left[\frac{3i}{2}
       \left\{\left(\frac{a}{b_0}\right)^2
                e^{-[2d/(n-2)+2]\kappa_\sigma \sigma}
\right\}^{-s+5/2}
       \int_{V_1}dz\;D_V(z)\:\cot(\pi z)
       \left(z^2-A_L^2\right)^{-s+5/2}
       \Gamma\left(s-\frac{5}{2}\right)\right.\nonumber\\
       & + &\left. i\left\{\left(\frac{a}{b_0}\right)^2
               e^{-[2d/(n-2)+2]\kappa_\sigma \sigma}
\right\}^{-s+3/2}
       \int_{V_1}dz\;D_V(z)\:\cot(\pi z)
       \left(z^2-A_L^2\right)^{-s+3/2}
        \Gamma\left(s-\frac{3}{2}\right)\right],
        \label{eq;c-6}
\end{eqnarray}
\end{widetext}
where $A^2_L$ is given by

%
\begin{equation}
A_L^2=D^2
      -(N^2+1)\left(\frac{b_0}{a}\right)^2
              e^{[2d/(n-2)+2]\kappa_\sigma \sigma}.
       \label{eq;c-7}
\end{equation}
and
the contour $V_1$ in the complex plane is showed 
in Fig.\ref{fig:v1}. 
Note that there are two branch points $z=\pm A_L$ in the integration.

Define the function $\tilde{\zeta}_k(s)$;
%
\begin{eqnarray}
\tilde{\zeta}_k(s)&=&
        \frac{\pi^{3/2}\Gamma\left(s-\frac{k}{2}\right)a^{2s}}
       {2^{2n-1}\left\{\Gamma(2N+1)\right\}^2\Gamma(s)}
         \nonumber\\
       & &\hspace{-1cm}\times \left[\frac{i}{2}
       \left\{\left(\frac{a}{b_0}\right)^2
               e^{-[2d/(n-2)+2]\kappa_\sigma \sigma}\right\}^{-s+k/2}
       \right.\nonumber\\ 
       & &\hspace{-1cm}\times\left.\int_{C_1}dz\;D_V(z)\:\cot(\pi z)
       \left(z^2-A_L^2\right)^{-s+k/2}\right].
         \nonumber\\
        \label{eq;c-8}
\end{eqnarray}
$\zeta_V(s)$ is then expressed by 
%
\begin{equation}
\zeta_V(s)=3\:\tilde{\zeta}_5(s)+8\:\tilde{\zeta}_3(s).
       \label{eq;c-9}
\end{equation}
 Now we move the
contour $V_1$ to the parallel line along the imaginary axis in order
to account the poles for $\cot(\pi z)$ in (\ref{eq;c-8}) (See Fig.\ref{fig:v1}).
The contour 
$V_2$ is replaced with  lines passing just above the cuts associated with $z=\pm A_L$.
$\zeta(s)$ is then given by 
%
\begin{eqnarray}
\tilde{\zeta}_k(s)&=&
       \frac{\pi^{3/2}a^{2s} }{2^{2n-1}
      \left\{\Gamma(2N+1)\right\}^2\Gamma(s)}
               \sin\left\{\pi\left(s-\frac{k}{2}\right)\right\}
         \nonumber\\
      & &\hspace{-1cm}\times \left\{\left(\frac{a}{b_0}\right)^2
               e^{-[2d/(n-2)+2]\kappa_\sigma \sigma}\right\}^{-s+k/2}
            \nonumber\\
      & &\hspace{-1cm}\times \left\{\int^{\infty}_0dx\;\:D_V(ix)
      \left(x^2+A_L^2\right)^{-s}\coth(\pi x)
           \right. \nonumber\\
      & &\hspace{-1cm}-\left.\int^{A_L}_0
      dx\:\cot(\pi x)\;D_V(x)\left(A_L^2-x^2\right)^{-s+k/2}
      \right\},
       \nonumber\\
       \label{eq;c-10}
\end{eqnarray}
where $D_V(ix)$ is the polynomial with coefficients $r_{Nk}$;
%
\begin{eqnarray}
D_V(ix)&=&(-1)^{D}\frac{2x^2}{(2D)!}
        \left\{x^2+\left(D-1\right)^2\right\}\cdots
        \left\{x^2+1\right\}\nonumber\\
       &\equiv&(-1)^{D}\sum^{D-1}_{p=0}\;r_{Np}\;x^{2p+2}.
       \label{eq;c-11}
\end{eqnarray}
The first term in Eq.(\ref{eq;c-8}) comes from the integral 
along the imaginary
axis and the  second  term in Eq.(\ref{eq;c-8}) 
is contribution from the
contours along the cuts of on the real axis, respectively.
Substituting  the relation 
%
\begin{equation}
\coth(\pi x)=1+\frac{2}{e^{2\pi x}-1},
       \label{eq;c-12}
\end{equation}
into the first term of (\ref{eq;c-10}), 
the function $\tilde{\zeta}_k(s)$ is finally given by  
%
\begin{widetext}
\begin{eqnarray}
\tilde{\zeta}_k(s)&=&\frac{\pi^{3/2}a^{2s} }{2^{2n-1}
      \left\{\Gamma(2N+1)\right\}^2\Gamma(s)}
    \sin\left\{\pi\left(s-\frac{k}{2}\right)\right\}
       \left\{\left(\frac{a}{b_0}\right)^2
                e^{-[2d/(n-2)+2]\kappa_\sigma \sigma}
      \right\}^{-s+k/2}\nonumber\\
      &\times&
      \left\{\frac{1}{2}\sum^{D-1}_{p=0}r_{Np}
      \left(A_L^2\right)^{p+4-s}\frac{\Gamma\left(p+\frac{3}{2}\right)
      \Gamma(s-p-4)}
      {\Gamma\left(s-\frac{k}{2}\right)}\right.
      \nonumber\\
    &-&\left.\int^{\infty}_0 dx\;D_V(ix)
       \left(x^2+A_L^2\right)^{-s+k/2}
       \frac{2}{e^{2\pi x}-1}
        +\int^{A_L}_0 dx D_V(x)
        \left(A_L^2-x^2\right)^{-s+k/2}\cot(\pi x)\right\}. 
        \nonumber\\
         \label{eq;c-13}
\end{eqnarray}
\end{widetext}

\end{document}